\documentclass[aps, prd, superscriptaddress, preprintnumbers]{revtex4}

\usepackage{graphicx,amsfonts,amsmath,amssymb,amstext}
\usepackage{float,wrapfig}
\usepackage{subfigure, psfrag}
\usepackage{dsfont}
\usepackage{color}
\usepackage{bm}

\newcommand{\be}{\begin{equation}}
\newcommand{\ee}{\end{equation}}
\newcommand{\ba}{\begin{eqnarray}}
\newcommand{\ea}{\end{eqnarray}}

\definecolor{purple}{rgb}{0.8,0,0.6}
\definecolor{darkgreen}{rgb}{0.00,0.6,0.00}

\begin{document}

\title{Non-Abelian properties of electron wave packets in Dirac semimetals A$_3$Bi (A=Na,K,Rb)}
\date{July 30, 2018}

\author{E.~V.~Gorbar}
\affiliation{Department of Physics, Taras Shevchenko National Kiev University, Kiev, 03680, Ukraine}
\affiliation{Bogolyubov Institute for Theoretical Physics, Kiev, 03680, Ukraine}

\author{V.~A.~Miransky}
\affiliation{Department of Applied Mathematics, Western University, London, Ontario, Canada N6A 5B7}

\author{I.~A.~Shovkovy}
\affiliation{College of Integrative Sciences and Arts, Arizona State University, Mesa, Arizona 85212, USA}
\affiliation{Department of Physics, Arizona State University, Tempe, Arizona 85287, USA}

\author{P.~O.~Sukhachov}
\affiliation{Department of Applied Mathematics, Western University, London, Ontario, Canada N6A 5B7}

\begin{abstract}
The motion of electron wave packets in the Dirac semimetals $\mathrm{A_3Bi}$ (A=Na,K,Rb) is studied in
a semiclassical approximation. Because of the two-fold degeneracy of the Dirac points and a momentum-dependent
gap term in the low-energy Hamiltonian, the associated Berry curvature is non-Abelian.
In the presence of background electromagnetic fields, such a Berry curvature leads to a splitting of trajectories
for the wave packets that originate from different Dirac points and chiral sectors. The nature of
the splitting strongly depends on the background fields as well as the initial chiral composition of the
wave packets. In parallel electric and magnetic fields, while a well-pronounced valley splitting is
achieved for any chirality composition, the chirality separation takes place predominantly for the initially polarized states.
On the other hand, in perpendicular electric and magnetic fields, there are clear deviations from the
conventional Abelian trajectories, albeit without a well-pronounced valley splitting.
\end{abstract}

\maketitle

\section{Introduction}
\label{sec:Introduction}

Dirac and Weyl semimetals are condensed matter materials whose low-energy excitations are described by the
Dirac and Weyl equations, respectively. Generically, the corresponding materials have a band structure
where the valence and conduction bands touch at isolated points (i.e., the Dirac points and the Weyl nodes, respectively).
Theoretically, $\mathrm{A_3Bi}$
(A=Na,K,Rb) and $\mathrm{Cd_3As_2}$ were the first compounds predicted to be Dirac semimetals with topologically
protected Dirac points \cite{Fang,WangWeng}. The existence of Dirac points in $\mathrm{Cd_3As_2}$ and
$\mathrm{Na_3Bi}$ was soon confirmed experimentally via the angle-resolved photoemission spectroscopy (ARPES)
in Refs.~\cite{Borisenko,Neupane,Liu}. Weyl semimetals were first predicted in pyrochlore iridates \cite{Savrasov},
but they were discovered experimentally in TaAs, TaP, NbAs, and NbP
\cite{Tong,Bian,Qian,Long,Xu-Hasan:TaP,Xu-Hasan:NbAs,Xu-Feng:NbP,Shekhar-Nayak:2015,Wang-Zheng:2015,Zhang-Xu:2015}
(for recent reviews, see Refs.~\cite{Hasan-Huang:2017-Rev,Yan-Felser:2017-Rev,Armitage-Vishwanath:2017-Rev}).

As it is well understood now, Weyl semimetals represent a topologically nontrivial phase of matter. Indeed, Weyl nodes are the monopoles of
the Berry curvature~\cite{Berry:1984} whose topological charges are directly connected with their chirality.
According to the Nielsen--Ninomiya theorem~\cite{Nielsen-Ninomiya-1,Nielsen-Ninomiya-2}, Weyl nodes
in crystals always come in pairs of opposite chirality. The corresponding nodes are separated in
momentum and/or energy. Such a nodal structure is also responsible for the existence of topologically
protected surface states, known as the Fermi arcs~\cite{Savrasov,Aji,Haldane}.

Unlike the Weyl nodes, the Dirac points are usually assumed to be topologically trivial because they
are composed from pairs of overlapping Weyl nodes of opposite chirality. By using numerical calculations
\cite{WangWeng,Fang}, however, it was found that the Dirac semimetals $\mathrm{Cd_3As_2}$ and
$\mathrm{A_3Bi}$ (A=Na,K,Rb) possess the Fermi arcs too. This was later confirmed experimentally
by the ARPES data \cite{Xu-Hasan:2015} and the observation of special surface-bulk quantum oscillations in
transport measurements \cite{Potter-Vishwanath:2014,Moll:2016}. It was argued
\cite{Yang-Nagaosa:2014,Gorbar:2014sja,Gorbar:2015waa,Yang-Furusaki:2015,Fang-Fu:2015,Kobayashi-Sato:2015,Burkov-Kim:2015} that the physical
reason for the nontrivial topological properties of $\mathrm{A_3Bi}$ (A=Na,K,Rb) is a
$Z_2$ symmetry that such materials possess. In the classification scheme proposed in Ref.~\cite{Yang-Nagaosa:2014},
such Dirac semimetals belong to the second class in which pairs of Dirac points are created by the
inversion of two bands. This is in contrast to the Dirac semimetals in the first class that possess a single
Dirac point at a time-reversal (TR) invariant momentum. As noted in Ref.~\cite{Burkov-Kim:2015}, the
presence of the $Z_2$ symmetry leads to the $Z_2$ anomaly that could affect transport properties.
The latter were recently discussed in Ref.~\cite{Rogatko:2018moa} using the hydrodynamic description.
A complementary view at the $Z_2$ symmetry in $\mathrm{A_3Bi}$ (A=Na,K,Rb) was presented in Refs.~\cite{Gorbar:2014sja,Gorbar:2015waa},
where we argued that these compounds are, in fact, hidden $Z_2$ Weyl semimetals. The discrete
symmetry of the low-energy effective Hamiltonian allows one to split all quasiparticle states into two
separate sectors, each describing a Weyl semimetal with a pair of Weyl nodes and a broken TR
symmetry. Since the $Z_2$ symmetry interchanges states from these two sectors, the TR symmetry
is preserved in the complete theory.

The degeneracy of opposite chirality states in the Dirac semimetals and the presence of the $Z_2$ symmetry
are expected to have profound consequences. The fact that the Berry curvature becomes a matrix with a
non-Abelian structure \cite{Wilczek:1984dh} could manifest itself, for example, in unusual transport properties
of the Dirac semimetals. The latter could be studied, for example, by employing the chiral kinetic theory
\cite{Son:2012wh,Stephanov:2012ki,Chen:2014cla,Manuel:2014dza} generalized to the case of
degenerate states \cite{Shindou:2005vfm,Culcer-Niu:2005,Chang:2008zza,Xiao:2009rm}.

The main motivation for this study is to investigate how the momentum-dependent gap term and the non-Abelian nature of the Berry curvature affect the quasiclassical properties of electron wave packets
in $Z_2$ Weyl semimetals. In particular, we consider the propagation of wave packets in external electric and magnetic fields.
Note that, in the absence of the
non-Abelian corrections to the Berry curvature, the semiclassical motion of chiral quasiparticles was
already considered in Ref.~\cite{Gorbar:2017dtp}, where the (pseudo-)magnetic lens was proposed.
It was found that while the primary contribution to the spatial splitting of quasiparticles of different chirality is related to the interplay of magnetic and strain-induced pseudomagnetic fields, the Abelian Berry curvature plays also an important, albeit auxiliary, role.
In this study, we investigate how the trajectories of the
wave packets change due to the presence of the off-diagonal gap term and the non-Abelian nature of
the Berry curvature. Of particular interest is the question as to whether the splitting of the wave packets
from different Dirac points (or, equivalently, valleys) and different chiral sectors can be achieved
without a background pseudomagnetic field.

The paper is organized as follows. In Sec.~\ref{sec:Model}, the low-energy effective model of the
Dirac semimetals $\mathrm{A_3Bi}$ (A=Na,K,Rb) and its linearized version are introduced.
We present the semiclassical equations of motion with the non-Abelian corrections in Sec.~\ref{sec:wavepacket-and-eqs}. The motion of the
electron wave packets in external electric and magnetic fields is investigated in Sec.~\ref{sec:trajectories-pm-DP}.
The results are discussed and summarized in Sec.~\ref{sec:Summary}. The expressions for the Berry connection, the Berry curvature,
and the magnetic moment of wave packets are given in Appendix \ref{sec:app-exp-expressions-lin-alpha}.

\section{Model}
\label{sec:Model}

In this section, we describe the low-energy model of the Dirac semimetals $\mathrm{A_3Bi}$ (A=Na,K,Rb)
as well as its linearized version and underlying symmetries. The corresponding quasiparticle Hamiltonian
derived in Ref.~\cite{Fang} reads as
\begin{equation}
\label{low-energy-Hamiltonian}
H(\mathbf{k}) = \epsilon_0(\mathbf{k}) I_4 + H_{4\times 4},
\end{equation}
where $I_4$ is the $4\times 4$ unit matrix, $\epsilon_0(\mathbf{k}) = C_0 + C_1k_z^2+C_2k_{\perp}^2$, $k_{\perp}=\sqrt{k_x^2+k_y^2}$, and
\begin{equation}
\label{low-energy-Hamiltonian4x4}
H_{4\times 4} =
\left( \begin{array}{cccc}
                   M(\mathbf{k}) & v_Fk_+ & 0 & \Delta^{*}(\mathbf{k}) \\
                   v_Fk_- & -M(\mathbf{k}) & \Delta^{*}(\mathbf{k}) & 0 \\
                   0 & \Delta(\mathbf{k}) & M(\mathbf{k}) & -v_Fk_- \\
                   \Delta(\mathbf{k}) & 0 & -v_Fk_+ & -M(\mathbf{k}) \\
        \end{array}
\right).
\end{equation}
The matrix Hamiltonian $H_{4\times 4}$ is naturally split into $2\times2$ blocks. The diagonal blocks are defined in terms of the
quadratic function $M(\mathbf{k}) = M_0 - M_1 k_z^2-M_2k_{\perp}^2$ and $v_Fk_{\pm}$, where $k_{\pm} = k_x\pm ik_y$.
The off-diagonal blocks are determined by the function $\Delta(\mathbf{k}) = \alpha k_zk_{+}^2$ that
plays a crucial role in this study and whose physical meaning will be discussed later.

The numerical values of parameters in Hamiltonian (\ref{low-energy-Hamiltonian}) can be determined by fitting the energy
spectrum obtained by the first-principles calculations \cite{Fang} and equal
\begin{equation}
\label{model-parameters}
\begin{array}{lll}
 C_0 = -0.06382~\mbox{eV},\qquad
& C_1 = 8.7536~\mbox{eV\,\AA}^2,\qquad
& C_2 = -8.4008~\mbox{eV\,\AA}^2,\\
 M_0=-0.08686~\mbox{eV},\quad
& M_1=-10.6424~\mbox{eV\,\AA}^2,\qquad
& M_2=-10.3610~\mbox{eV\,\AA}^2,\\
 v_F=2.4598~\mbox{eV\,\AA}.
\end{array}
\end{equation}
Note that the Fermi velocity $v_F$ is given in energy units. Since no specific value for $\alpha$, which determines the
magnitude of the off-diagonal terms, was quoted in Ref.~\cite{Fang}, we will treat it as a free, albeit small, parameter
below. In addition to the model parameters in Eq.~(\ref{model-parameters}), we will also need the transport scattering time $\tau$.
For the purposes of this study, we use $\tau \approx10^{-10}~\mbox{s}$, which is an
estimated value of the scattering time in $\mathrm{Cd_3As_2}$ \cite{Liang-Ong:2015}.

\begin{figure}[ht]
\begin{center}
\includegraphics[width=0.45\textwidth]{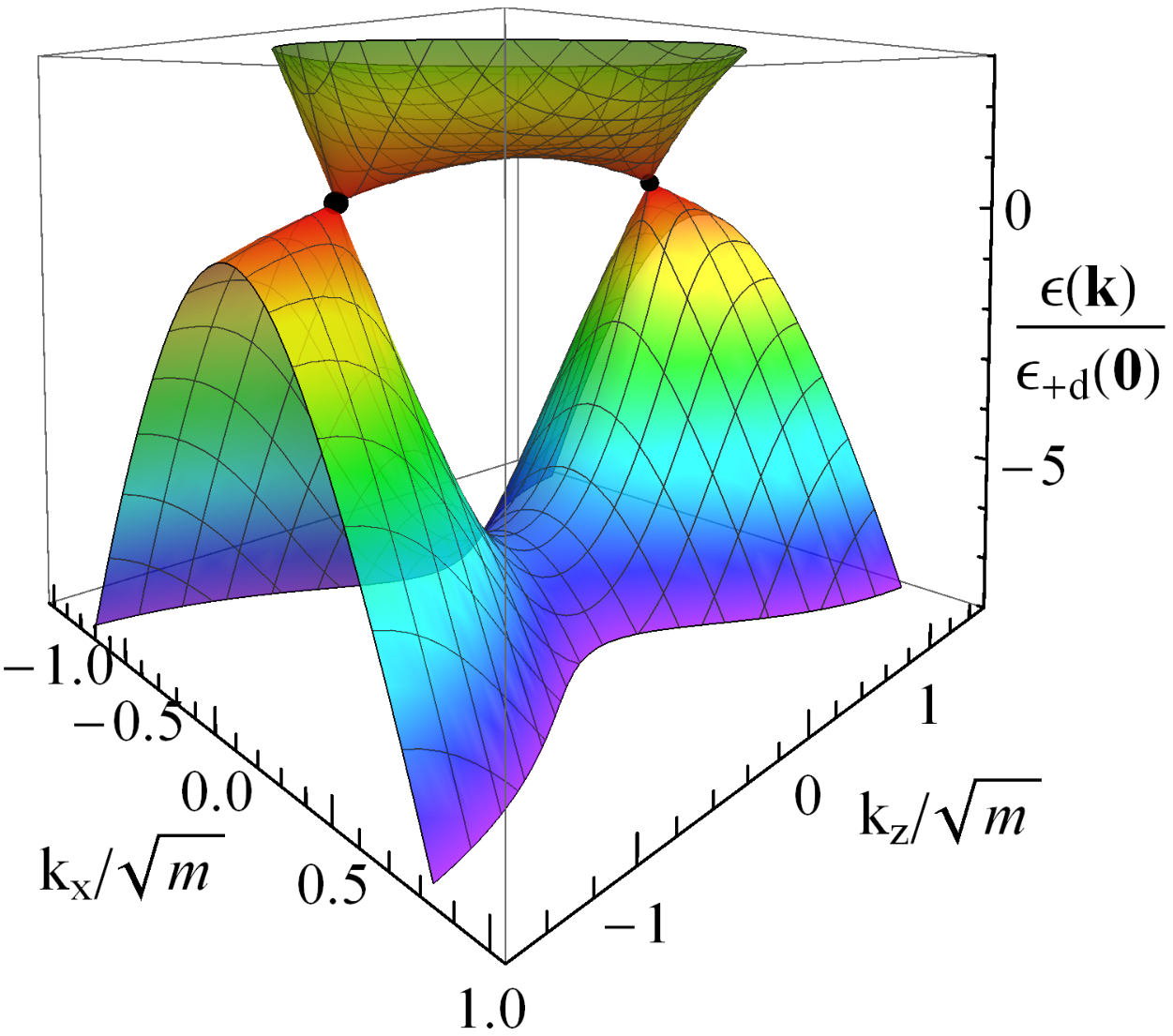}\hfill
\includegraphics[width=0.45\textwidth]{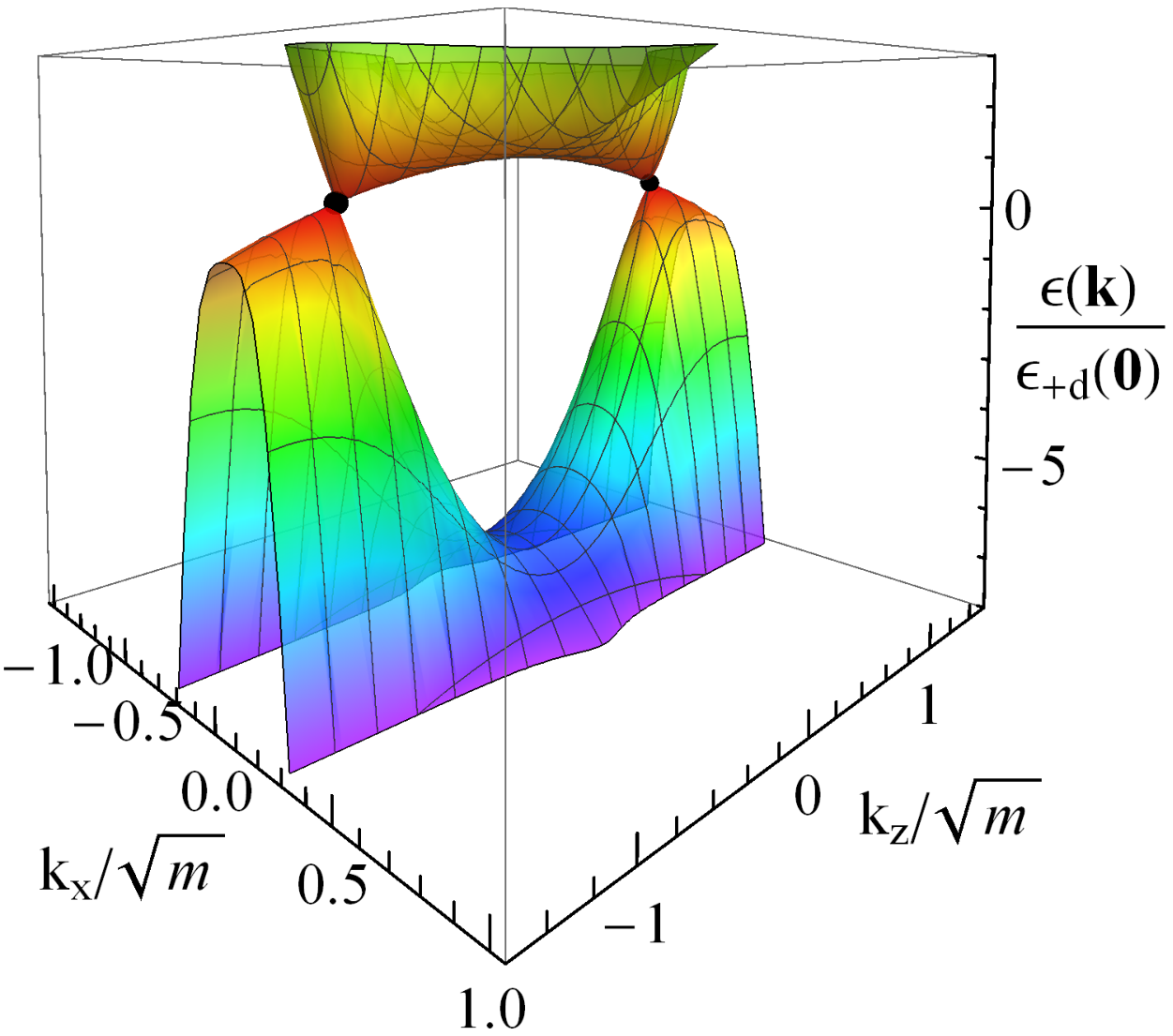}\hfill
\caption{The energy spectrum of the low-energy model (\ref{low-energy-Hamiltonian})
at $\alpha=0$ (left panel) and $\alpha=10\alpha^{*}$ (right panel), where $\alpha^{*}$
is the characteristic value defined in Eq.~(\ref{model-alpha-def-2}) and $\epsilon_{+d}(0)$ is defined in Eq.~(\ref{model-epsk0}).}
\label{fig:model-energy-full}
\end{center}
\end{figure}

The energy eigenvalues of Hamiltonian (\ref{low-energy-Hamiltonian}) are given by the following expression:
\begin{equation}
\label{energy-dispersion}
\epsilon(\mathbf{k})=\epsilon_0(\mathbf{k}) \pm \sqrt{M^2(\mathbf{k})+v_F^2k_{\perp}^2+|\Delta(\mathbf{k})|^2}.
\end{equation}
As is clear, a nonzero $\epsilon_0(\mathbf{k})$ introduces an asymmetry between the positive (electrons) and
negative (holes) energy branches and, consequently, breaks the particle-hole symmetry. The square root term
vanishes at the two Dirac points, $\mathbf{k}^{(\pm)}_0=\left(0, 0,  \pm \sqrt{m}\right)$, where $\sqrt{m}= \sqrt{M_0/M_1}$.
By using the low-energy parameters in Eq.~(\ref{model-parameters}), we find that $\sqrt{m}\approx 0.0903~\mbox{\AA}^{-1}$.
The latter defines the characteristic momentum scale in the low-energy Hamiltonian. Therefore, by equating the last two terms
under the square root in Eq.~(\ref{energy-dispersion}) and setting $k_z=k_{\perp}=\sqrt{m}$, we can estimate the characteristic value
of parameter $\alpha$, i.e.,
\begin{equation}
\label{model-alpha-def-2}
\alpha^{*} = \frac{v_F}{m} \approx 301.384~\mbox{eV}\mbox{\AA}^3.
\end{equation}
In order to get a better insight into the role of the off-diagonal term in the low-energy Hamiltonian, we plot the
corresponding energy spectra for $\alpha=0$ and $\alpha=10\alpha^{*}$ in the two panels of
Fig.~\ref{fig:model-energy-full}. As expected from Eq.~(\ref{energy-dispersion}), there are two Dirac points
well separated in $k_z$. The term $\Delta(\mathbf{k})$ plays the role of a momentum-dependent gap function that
mixes eigenstates of opposite chirality. While $\Delta(\mathbf{k})$ can profoundly change the spectrum of quasiparticles
for sufficiently large $k_{\perp}$, it vanishes at the Dirac points. Thus, the upper and lower $2 \times 2$ blocks of
Hamiltonian~(\ref{low-energy-Hamiltonian4x4}) still describe quasiparticle states of opposite chirality in a sufficiently
close vicinity of the Dirac points, although, strictly speaking, the notion of chirality is rigorous only at $\alpha=0$.

As discussed in detail in Refs.~\cite{Gorbar:2014sja,Gorbar:2015waa}, the actual form of function $\Delta(\mathbf{k})$
is consistent with the discrete $Z_2$ symmetry, implying that the Dirac semimetals $\mathrm{A_3Bi}$ (A=Na,K,Rb)
are effectively the hidden $Z_2$ Weyl semimetals. The quasiparticle states of these materials can be naturally split by using the ud (up-down)
symmetry \cite{Gorbar:2014sja}
\begin{equation}
\label{model-ud-parity}
U_{\chi}=\Pi_{k_z\to-k_z} \left(
                            \begin{array}{cc}
                              I_2 & 0 \\
                              0 & -I_2 \\
                            \end{array}
                          \right),
\end{equation}
where $\Pi_{k_z\to-k_z}$ is the operator that changes the sign of the $z$ component of momentum and $I_2$ is the
$2\times2$ unit matrix. The TR symmetry is broken in each of the $Z_2$ sectors that signifies the presence of the Weyl semimetal phase with the Weyl nodes separated by $2\sqrt{m}$.
Since the chirality of the nodes in different $Z_2$ sectors is opposite, the
complete model preserves the TR symmetry and has two Dirac points.

In order to simplify the calculations, we will omit the term $\epsilon_0(\mathbf{k})$ and linearize
Hamiltonian (\ref{low-energy-Hamiltonian}) in the vicinity of the Dirac points $\mathbf{k}^{(\pm)}_0$.
By expanding $M(\mathbf{k})$ up to the linear order in deviations $\mathbf{\delta{k}}=\mathbf{k}-\mathbf{k}^{(\pm)}_0$
and performing the unitary transformation with $U_x=\mbox{diag}(\sigma_x, I_2)$, we obtain
\begin{equation}
\label{model-Hamiltonian-canonical-plus}
H_{\rm lin}^{(+)}(\tilde{\mathbf{k}})=\left(
         \begin{array}{cc}
          v_F\left(\tilde{k}_x\sigma_x+\tilde{k}_y\sigma_y-\tilde{k}_z\sigma_z\right) & \alpha\left(\sqrt{m}+\tilde{k}_z\right)\tilde{k}_{-}^2 \\
          \alpha\left(\sqrt{m}+\tilde{k}_z\right)\tilde{k}_{+}^2 & -v_F\left(\tilde{k}_x\sigma_x+\tilde{k}_y\sigma_y-\tilde{k}_z\sigma_z\right) \\
         \end{array}
       \right)
\end{equation}
in the vicinity of the Dirac point at $\mathbf{k}_0^{(+)}$ and
\begin{equation}
\label{model-Hamiltonian-canonical-minus}
H_{\rm lin}^{(-)}(\tilde{\mathbf{k}})=\left(
         \begin{array}{cc}
          v_F\,(\tilde{\mathbf{k}}\cdot\bm{\sigma}) & -\alpha\left(\sqrt{m}-\tilde{k}_z\right)\tilde{k}_{-}^2 \\
          -\alpha\left(\sqrt{m}-\tilde{k}_z\right)\tilde{k}_{+}^2 & -v_F\,(\tilde{\mathbf{k}}\cdot\bm{\sigma}) \\
         \end{array}
       \right)
\end{equation}
in the vicinity of the Dirac point at $\mathbf{k}_0^{(-)}$. Here $\bm{\sigma}$ are the Pauli matrices and
$\tilde{\mathbf{k}} = ( k_x,k_y,2\sqrt{M_0M_1}\delta k_z/v_F )$. In the model at hand $2\sqrt{M_0M_1}
\approx0.78v_F$ and, consequently, the quasiparticle energy spectra near the Dirac points can be approximately considered as isotropic $\tilde{\mathbf{k}}=(k_x,k_y,\delta k_z )$.
The corresponding positive branches of the energies for Hamiltonians (\ref{model-Hamiltonian-canonical-plus}) and (\ref{model-Hamiltonian-canonical-minus}) are given by
\begin{equation}
\label{energy-dispersion-lin}
\epsilon^{(\pm)}(\tilde{\mathbf{k}})= \sqrt{v_F^2\tilde{k}^2+\alpha^2
\left(\sqrt{m}\pm\tilde{k}_z\right)^2 k_{\perp}^4 },
\end{equation}
where the superscript labels the Dirac points at $\mathbf{k}_0^{(\pm)}$. Note that we keep the $\tilde{k}_z$
term in the off-diagonal components in Eqs.~(\ref{model-Hamiltonian-canonical-plus}) and
(\ref{model-Hamiltonian-canonical-minus}) because, as will be clear below, it is relevant for the Berry
connection and the magnetic moment, which contain derivatives with respect to the $z$ component of
momentum. Also, this term plays an important role in determining the wave packet velocity.

Obviously, the dynamics of quasiparticles can be reliably described in terms of the two independent linearized
Hamiltonians (\ref{model-Hamiltonian-canonical-plus}) and (\ref{model-Hamiltonian-canonical-minus}) only for
sufficiently small energies and deviations of momentum, $|\delta k_z|\lesssim\sqrt{m}$.
The constraint $|\delta k_z|\lesssim\sqrt{m}$ also ensures that the internode transitions are negligible.
In order to obtain the characteristic energy scales of the low-energy region, we
calculate the height of the energy ``domes" in the full Hamiltonian (\ref{low-energy-Hamiltonian}) at
$\mathbf{k}=\mathbf{0}$, see also Fig.~\ref{fig:model-energy-full}. The corresponding values read as
\begin{equation}
\label{model-epsk0}
\epsilon_{+d}(0) = 23.0~\mbox{meV}, \qquad
\epsilon_{-d}(0) = -150.7~\mbox{meV}
\end{equation}
for positive and negative energies, respectively. Without the term $\epsilon_0(\mathbf{k})$, we have
\begin{equation}
\label{model-epsk0-pm}
\epsilon_{+d}(0)\big|_{\epsilon_0=0}=-\epsilon_{-d}(0)\big|_{\epsilon_0=0} = 86.9~\mbox{meV}.
\end{equation}
In order to simplify our notations, in the following we assume that the momentum $\mathbf{k}$ is measured
from the corresponding Dirac points, i.e., we replace $\delta k_z$ with $k_z$.

The energy spectrum (\ref{energy-dispersion-lin}) at each Dirac point is doubly degenerate in energy with the
corresponding wave functions given by
\begin{eqnarray}
\label{WPE-psi-def-p}
\psi_{+, \mathbf{k}}^{(\pm)} &=& \frac{v_F k_{\perp}}{\sqrt{\left[\epsilon^{(\pm)}(\mathbf{k}) \pm v_F k_z\right]^2+v_F^2k_{\perp}^2 +\alpha^2\left(k_z\pm\sqrt{m}\right)^2k_{\perp}^2}}
                                                    \left(
                                                    \begin{array}{c}
                                                      1 \\
                                                      \frac{\epsilon^{(\pm)}(\mathbf{k})\pm v_Fk_z}{v_F k_{-}} \\
                                                      0 \\
                                                      \frac{\alpha \left(k_z\pm\sqrt{m}\right)k_{+}^2}{v_Fk_{-}} \\
                                                    \end{array}
                                                  \right), \\
\label{WPE-psi-def-m}
                                                  \psi_{-, \mathbf{k}}^{(\pm)} &=&
                                                  \frac{v_F k_{\perp}}{\sqrt{\left[\epsilon^{(\pm)}(\mathbf{k}) \mp v_F k_z\right]^2+v_F^2k_{\perp}^2 +\alpha^2\left(k_z\pm\sqrt{m}\right)^2k_{\perp}^2}}
                                                  \left(
                                                    \begin{array}{c}
                                                      0 \\
                                                      \frac{\alpha \left(k_z\pm\sqrt{m}\right)k_{-}^2}{v_Fk_{-}} \\
                                                      1 \\
                                                      -\frac{\epsilon^{(\pm)}(\mathbf{k}) \mp v_Fk_z}{v_F k_{-}} \\
                                                    \end{array}
                                                  \right).
\end{eqnarray}
Here the upper index corresponds to the Dirac points at $\mathbf{k}_0^{(\pm)}$, which are described by the linearized Hamiltonians
(\ref{model-Hamiltonian-canonical-plus}) and (\ref{model-Hamiltonian-canonical-minus}), respectively. As we will see below,
this degeneracy is responsible for the non-Abelian nature of the Berry curvature in the Dirac semimetals $\mathrm{A_3Bi}$
(A=Na,K,Rb). It also implies that the semiclassical equations of motion for a degenerate case
\cite{Shindou:2005vfm,Culcer-Niu:2005,Chang:2008zza,Xiao:2009rm} should be used.
In order to describe this degeneracy, we introduce the following transformation
that can be viewed as an analog of the discrete chiral symmetry:
\begin{equation}
\label{model-ud-parity-D}
\Gamma_5=\Pi_{\alpha\to-\alpha} \left(
                            \begin{array}{cc}
                              I_2 & 0 \\
                              0 & -I_2 \\
                            \end{array}
                          \right).
\end{equation}
Note that $\Gamma_5$ is not a true symmetry because it does not commute with the linearized Hamiltonians (\ref{model-Hamiltonian-canonical-plus}) and (\ref{model-Hamiltonian-canonical-minus}).
The wave functions
$\psi_{+, \mathbf{k}}^{(\pm)}$ and $\psi_{-, \mathbf{k}}^{(\pm)}$ are the eigenstates of $\Gamma_5$, i.e., $\Gamma_5\psi_{+, \mathbf{k}}^{(\pm)}=\psi_{+, \mathbf{k}}^{(\pm)}$ and $\Gamma_5\psi_{-, \mathbf{k}}^{(\pm)}=-\psi_{-, \mathbf{k}}^{(\pm)}$, that describe the states with the positive and negative chirality, respectively, in the limit $\alpha \to 0$.
In addition, the positive branch of the band energy for the linearized Hamiltonians (\ref{model-Hamiltonian-canonical-plus}) and
(\ref{model-Hamiltonian-canonical-minus}) is
\begin{equation}
\label{WPE-eps-def}
\epsilon^{(\pm)}(\mathbf{k}) =  \sqrt{v_F^2 k^2 +\alpha^2 \left(\sqrt{m}\pm k_z\right)^2 k_{\perp}^4}\approx v_F k + O(\alpha^2).
\end{equation}
Henceforth, we will consider only the electron wave packets with positive energies.

\section{Non-Abelian corrections to the equations of motion}
\label{sec:wavepacket-and-eqs}

In this section, we consider the electron wave packets in the $Z_2$ Weyl semimetals and present the corresponding equations of
motion. Since we treat the Dirac points as independent and, consequently, there are no internode mixing terms, the superscript $\pm$
for all quantities will be omitted in this section. An electron wave packet centered at $\mathbf{r}(t)$ and $\mathbf{q}(t)$ is
defined as a superposition of the Bloch states $\phi_{n, \mathbf{k}}=e^{i\mathbf{k}\mathbf{r}}\psi_{n,\mathbf{k}}$, i.e.,
\begin{equation}
\label{WPE-W-def}
W = \sum_{n=\pm} \int \frac{d\mathbf{k}}{(2\pi)^3} a(t,\mathbf{k}) \eta_n(t,\mathbf{k}) \phi_{n, \mathbf{k}}.
\end{equation}
Here $n=\pm$ denotes the degenerate chiral states and $a(t,\mathbf{k})$ is a normalized distribution
centered at $\mathbf{r}(t)$ and $\mathbf{q}(t)$. Finally, $\eta_n(t,\mathbf{k})$ denotes the partial
contributions or weights of the degenerate states, satisfying the normalization condition
$\sum_{n=\pm}|\eta_n(t,\mathbf{k})|^2=1$.

As is well known, the nontrivial topological properties of Weyl semimetals are captured by the monopole-like
Berry curvature \cite{Berry:1984} at the Weyl nodes. Because of the additional $\Gamma_5$-chirality degree
of freedom at each Dirac point in the $\mathrm{A_3Bi}$ (A=Na,K,Rb) semimetals, the corresponding Berry
connection is a $2\times 2$ matrix. Its elements are defined by
\begin{equation}
\label{WPE-Berry-connection-def}
\mathbf{A}_{nm}(\mathbf{q}) = -\frac{i}{2}\left(\psi_{n, \mathbf{q}}^{\dag} \partial_{\mathbf{q}} \psi_{m, \mathbf{q}} - \psi_{m, \mathbf{q}}^{\dag} \partial_{\mathbf{q}}
\psi_{n, \mathbf{q}}\right).
\end{equation}
The explicit expressions for $\mathbf{A}_{nm}$ are given by Eqs.~(\ref{exp-expressions-lin-alpha-A++})--(\ref{exp-expressions-lin-alpha-A--})
in Appendix \ref{sec:app-exp-expressions-lin-alpha}. (Note that the off-diagonal components of $\mathbf{A}_{nm}$ vanish when $\alpha=0$.)
The Berry curvature has a non-Abelian structure, i.e.,
\begin{equation}
\label{WPE-Berry-curvature-def}
\bm{\Omega}_{nm}(\mathbf{q}) = -\frac{i}{\hbar}\sum_{l=\pm}\left[(D_{\mathbf{q}})_{nl} \times (D_{\mathbf{q}})_{lm} \right]
=\frac{1}{\hbar}\left[\partial_{\mathbf{q}} \times \mathbf{A}_{nm}(\mathbf{q})\right] + \frac{i}{\hbar}\sum_{l=\pm}
\left[\mathbf{A}_{nl}(\mathbf{q})\times \mathbf{A}_{lm}(\mathbf{q})\right],
\end{equation}
where $(D_{\mathbf{q}})_{nl} = \partial_{\mathbf{q}} \delta_{nl} +i \mathbf{A}_{nl}(\mathbf{q})$ is the covariant derivative.
The components of $\bm{\Omega}_{nm}(\mathbf{q})$ are given by
Eqs.~(\ref{exp-expressions-lin-alpha-Omega++})--(\ref{exp-expressions-lin-alpha-Omega--})
in Appendix \ref{sec:app-exp-expressions-lin-alpha}. The semiclassical Hamiltonian is defined by
\begin{equation}
\label{WPE-H-cal-def}
\mathcal{H}_{nm}(\mathbf{r},\mathbf{q}) = \left[\epsilon(\mathbf{q}) -e\varphi(\mathbf{r})\right]\delta_{nm}
+\left(\mathbf{M}_{nm}(\mathbf{q})\cdot\mathbf{B}\right)
\end{equation}
and contains the band, electrostatic, as well as the magnetization energy determined by the
magnetic moment of the wave packet, i.e.,
\begin{equation}
\label{WPE-magnetic-moment-def}
\mathbf{M}_{nm}(\mathbf{q}) = i\frac{e}{2\hbar c} \left[(\partial_{\mathbf{q}}\psi^{\dag}_{n, \mathbf{q}}) \times \left\{H(\mathbf{q}) -\epsilon(\mathbf{q}) I_2 \right\}(\partial_{\mathbf{q}}\psi_{m, \mathbf{q}})\right].
\end{equation}
Here $H(\mathbf{q})$ is given by $H_{\rm lin}^{(\pm)}(\mathbf{q})$ in Eqs.~(\ref{model-Hamiltonian-canonical-plus})
and (\ref{model-Hamiltonian-canonical-minus}) for the Dirac points at $\mathbf{k}_0^{(\pm)}$, respectively, and the
band energy $\epsilon(\mathbf{q})$ is defined by Eq.~(\ref{energy-dispersion-lin}). The explicit expressions for the
components of the magnetic moment (\ref{WPE-magnetic-moment-def}) are given by
Eqs.~(\ref{exp-expressions-lin-alpha-M++})--(\ref{exp-expressions-lin-alpha-M--}) in Appendix \ref{sec:app-exp-expressions-lin-alpha}.

The equations of motion for the non-Abelian wave packet in constant external electric $\mathbf{E}$ and magnetic $\mathbf{B}$
fields are given by \cite{Culcer-Niu:2005}
\begin{eqnarray}
\label{WPE-r-eq-def}
\dot{\mathbf{r}} &=& \mathbf{v}(\mathbf{q}) + \hbar \left[\dot{\mathbf{q}} \times \bm{\Omega}(\mathbf{q})\right],\\
\label{WPE-q-eq-def}
\hbar \dot{\mathbf{q}} &=& -e\mathbf{E} - \frac{e}{c}\left[\dot{\mathbf{r}}\times\mathbf{B}\right] - \frac{\hbar \mathbf{q}}{\tau},\\
\label{WPE-eta-eq-def}
i\hbar\, \dot{\eta}_n &=& \left[\left(\mathbf{M}_{nm}(\mathbf{q})\cdot\mathbf{B}\right)
+ \hbar \left(\dot{\mathbf{q}}\cdot\mathbf{A}_{nm}(\mathbf{q})\right)\right] \eta_{m},
\end{eqnarray}
where the wave packet's velocity is defined by
\begin{eqnarray}
\label{WPE-v-mean-def}
\mathbf{v}(\mathbf{q}) &=& \frac{1}{\hbar}\sum_{n,m,l=\pm} \eta^{\dag}_n \left[(D_{\mathbf{q}})_{nl}, \mathcal{H}_{lm}(\mathbf{r},\mathbf{q})\right] \eta_m = \frac{1}{\hbar} \partial_{\mathbf{q}}\epsilon(\mathbf{q})\nonumber\\
&+& \frac{1}{\hbar}\sum_{n,m,l=\pm} \eta^{\dag}_n \left\{ \delta_{ln}\left[\partial_{\mathbf{q}}\left(\mathbf{M}_{nm}(\mathbf{q})\cdot\mathbf{B}\right)\right] + i\left[\mathbf{A}_{nl}(\mathbf{q})\left(\mathbf{M}_{lm}(\mathbf{q})\cdot\mathbf{B}\right) -\left(\mathbf{M}_{nl}(\mathbf{q})\cdot\mathbf{B}\right)\mathbf{A}_{lm}(\mathbf{q})\right]\right\} \eta_m
\end{eqnarray}
and the Berry curvature reads
\begin{equation}
\label{WPE-Omega-mean-def}
\bm{\Omega}(\mathbf{q}) = \sum_{n,m=\pm} \eta^{\dag}_n \bm{\Omega}_{nm}(\mathbf{q}) \eta_m.
\end{equation}
It is worth noting that the non-Abelian equations of motion (\ref{WPE-r-eq-def})--(\ref{WPE-eta-eq-def}) differ from those for Abelian wave packets by the presence of an additional equation for the weights of the degenerate states $\eta_n$, i.e., Eq.~(\ref{WPE-eta-eq-def}).
Note also that a phenomenological dissipative term
$\hbar \mathbf{q}/\tau$ was introduced in Eq.~(\ref{WPE-q-eq-def}). Physically, it captures the effects of
scattering of the electrons on impurities, defects, and phonons in the relaxation time approximation.

The system of equations (\ref{WPE-r-eq-def})--(\ref{WPE-eta-eq-def}) can be rewritten in a more convenient form where all derivatives
are grouped on the left-hand sides, i.e.,
\begin{eqnarray}
\label{WPE-r-eq-exp}
\dot{\mathbf{r}} \left[1-\frac{e}{c} \left(\bm{\Omega}\cdot\mathbf{B}\right)\right]
&=& \mathbf{v} -e\left[\mathbf{E}\times\bm{\Omega}\right] -\frac{e}{c} \mathbf{B} \left(\bm{\Omega}\cdot\mathbf{v}\right) -\frac{\hbar\left[\mathbf{q}\times\bm{\Omega}\right]}{\tau},\\
\label{WPE-q-eq-exp}
\hbar \dot{\mathbf{q}} \left[1-\frac{e}{c} \left(\bm{\Omega}\cdot\mathbf{B}\right)\right]
&=& \mathbf{F},\\
\label{WPE-eta-eq-exp}
i\hbar\, \dot{\eta}_n \left[1-\frac{e}{c} \left(\bm{\Omega}\cdot\mathbf{B}\right)\right] &=& \sum_{m=\pm}\left\{\left(\mathbf{F}\cdot\mathbf{A}_{nm}\right)
+\left(\mathbf{M}_{nm}\cdot\mathbf{B}\right)\left[1-\frac{e}{c} \left(\bm{\Omega}\cdot\mathbf{B}\right)\right]\right\} \eta_{m}.
\end{eqnarray}
Here we used the following the short-hand notation:
\begin{equation}
\label{force}
\mathbf{F}=-e\mathbf{E} -\frac{e}{c} \left[\mathbf{v}\times \mathbf{B}\right] + \frac{e^2}{c} \bm{\Omega}\left(\mathbf{E}\cdot\mathbf{B}\right) -\frac{\hbar \mathbf{q}}{\tau} \left[1+\frac{e}{c} \left(\bm{\Omega}\cdot\mathbf{B}\right)\right] + \frac{e\hbar \bm{\Omega}}{c\tau} \left(\mathbf{q}\cdot\mathbf{B}\right).
\end{equation}
For simplicity of presentation, here we omitted the arguments of $\bm{\Omega}$, $\mathbf{v}$, $\mathbf{M}_{nm}$,
and $\mathbf{A}_{nm}$. As is easy to see, the presence of the non-Abelian corrections complicates significantly the
equations of motion. As a result, the latter can be solved only numerically. The corresponding solutions for the cases
of perpendicular and parallel electric and magnetic fields are discussed in the next section.

\section{Wavepackets motion}
\label{sec:trajectories-pm-DP}

As discussed in the previous section, the time evolution of the coordinates, momenta, and partial weights
of the wave packets is described by Eqs.~(\ref{WPE-r-eq-exp}), (\ref{WPE-q-eq-exp}), and (\ref{WPE-eta-eq-exp}).
The corresponding equations should be also supplemented by the initial conditions. In view of the translation
invariance of the problem, we can set without the loss of generality the initial coordinates of the wave packet
to be at the origin of the coordinate system, i.e.,
\begin{equation}
\label{trajectories-Ey-Bz-init-val-r}
\mathbf{r}(t=0)=\mathbf{0}.
\end{equation}
As for the initial value of the wave packet's momentum, it is convenient to match it with the steady-state
value determined by the electric field in the relaxation time approximation, i.e.,
\begin{equation}
\label{trajectories-Ey-Bz-init-val-q}
\mathbf{q}(t=0)= - \frac{e\tau \mathbf{E}}{\hbar}.
\end{equation}
Concerning the initial conditions for the partial weights $\eta_{\pm}$, it is natural to assume that the
wave packets are non-chiral with respect to the $\Gamma_5$ transformation (i.e., the probabilities to
find an electron in the positive and negative $\Gamma_5$-chirality states are equal), i.e.,
\begin{equation}
\label{trajectories-Ey-Bz-init-val-eta-1}
\eta_{+}(t=0)=  \eta_{-}(t=0)= \frac{1}{\sqrt{2}}.
\end{equation}
For the sake of completeness, however, in Sec.~\ref{sec:trajectories-pm-DP-exact-tau-2-polarization} we will also consider
the case of the initially polarized wave packets with
\begin{equation}
\label{trajectories-pm-DP-exact-tau-2-polarization-eta-1}
\eta_{+}(t=0) = 1, \qquad \eta_{-}(t=0) = 0
\end{equation}
and
\begin{equation}
\label{trajectories-pm-DP-exact-tau-2-polarization-eta-2}
\eta_{+}(t=0) = 0, \qquad \eta_{-}(t=0) = 1.
\end{equation}

Let us begin our consideration with the case when the background magnetic field is absent, $\mathbf{B}=\mathbf{0}$.
As is easy to see, the structure of the equations of motion (\ref{WPE-r-eq-exp}), (\ref{WPE-q-eq-exp}),
and (\ref{WPE-eta-eq-exp}) drastically simplifies, i.e.,
\begin{eqnarray}
\label{WPE-r-eq-exp-B=0}
\dot{\mathbf{r}} &=& \mathbf{v} -e\left[\mathbf{E}\times\bm{\Omega}\right] -\frac{\hbar\left[\mathbf{q}\times\bm{\Omega}\right]}{\tau},\\
\label{WPE-q-eq-exp-B=0}
\hbar \dot{\mathbf{q}} &=& -e\mathbf{E} -\frac{\hbar \mathbf{q}}{\tau},\\
\label{WPE-eta-eq-exp-B=0}
i\hbar\, \dot{\eta}_n  &=& -\sum_{m=\pm}\left(\left[e\mathbf{E} +\frac{\hbar \mathbf{q}}{\tau} \right]\cdot\mathbf{A}_{nm}\right)\eta_{m}.
\end{eqnarray}
For the initial conditions in Eqs.~(\ref{trajectories-Ey-Bz-init-val-r}) and (\ref{trajectories-Ey-Bz-init-val-q}),
we obtain the following analytical solution:
\begin{eqnarray}
\label{WPE-r-sol-B=0}
\mathbf{r}(t) &=& \mathbf{v} t,\\
\label{WPE-q-sol-B=0}
\mathbf{q}(t) &=& -\frac{\tau e\mathbf{E}}{\hbar},\\
\label{WPE-eta-sol-B=0}
\eta_n(t)  &=& \eta_n(0),
\end{eqnarray}
which describes the inertial motion of wave packets with no mixing of the $\Gamma_5$-chirality states.
It is worth noting that this result is valid for both full and linearized Hamiltonians given in Eq.~(\ref{low-energy-Hamiltonian})
as well as Eqs.~(\ref{model-Hamiltonian-canonical-plus}) and (\ref{model-Hamiltonian-canonical-minus}), respectively.

As we will see below, the dynamics of wave packets becomes considerably more
complicated when an external magnetic field is present. The cases of parallel and perpendicular electromagnetic
fields are studied in the next two subsections.

\subsection{Parallel electric and magnetic fields}
\label{sec:trajectories-pm-DP-exact-tau-2-Ey-By}

In this subsection, we study the motion of wave packets in parallel electric and magnetic fields when the initial
chiral weights are equal, i.e., $\eta_{+}(t=0)=\eta_{-}(t=0)=1/\sqrt{2}$. In order to solve the equations of motion
numerically, we set $\mathbf{E}=E\hat{\mathbf{y}}$ and $\mathbf{B}=B\hat{\mathbf{y}}$, where $E=200~\mbox{V/m}$ and $B=10~\mbox{G}$.

\begin{figure}[t]
\begin{center}
\includegraphics[width=0.45\textwidth]{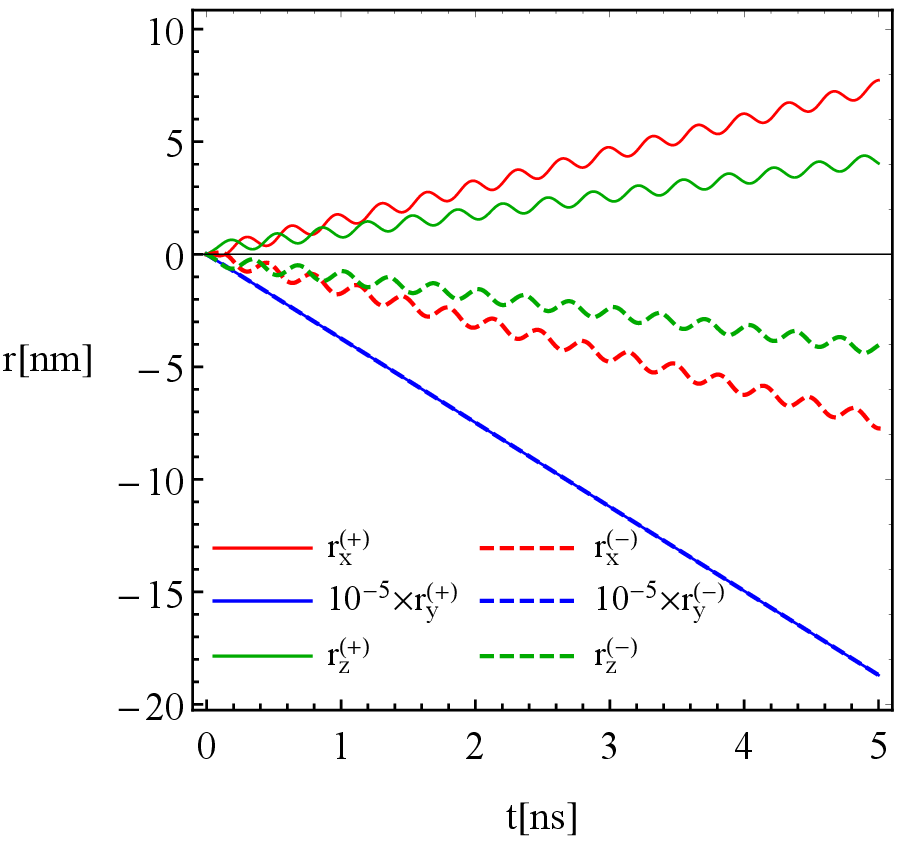}\hfill
\includegraphics[width=0.45\textwidth]{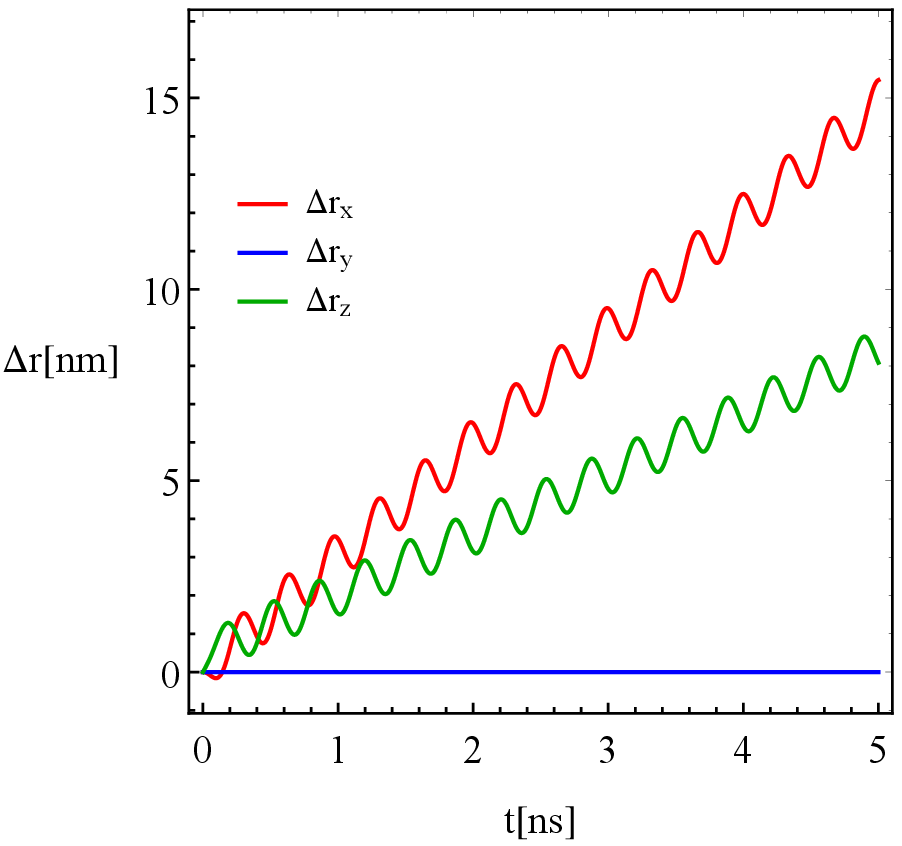}\\
\caption{The positions $\mathbf{r}^{(\pm)}$ of the wave packets as a function of time $t$ (left panel) and the splitting
$\Delta \mathbf{r}=\mathbf{r}^{(+)}-\mathbf{r}^{(-)}$ between the coordinates of the wave packets from different Dirac
points (right panel). The red, blue, and green lines correspond to the $x$, $y$, and $z$ components, respectively.
The solid and dashed lines represent the results for the wave packets described by Hamiltonians
(\ref{model-Hamiltonian-canonical-plus}) and (\ref{model-Hamiltonian-canonical-minus}), respectively.
We used $\alpha=0.5\alpha^{*}$, $\mathbf{E}=E\hat{\mathbf{y}}$, and $\mathbf{B}=B\hat{\mathbf{y}}$, where
$E=200~\mbox{V/m}$ and $B=10~\mbox{G}$.}
\label{fig:trajectories-pm-DP-exact-tau-2-Ey-By-r}
\end{center}
\end{figure}

The position vectors $\mathbf{r}^{(\pm)}$ of the wave packets from different valleys are shown
in Fig.~\ref{fig:trajectories-pm-DP-exact-tau-2-Ey-By-r}. We find that the momentum-dependent chirality-mixing
leads to a noticeable splitting of the wave packets in the $x$ and $z$ directions that increases with time.
At the same time, the splitting in the $y$ direction is negligible. We also find that the non-Abelian terms give rise
to periodic oscillations of the wave packets around their overall linear trajectories. As we argue below, the physical
origin of such oscillations is connected with the precession of the magnetic moment.

Further, we find that the momenta of wave packets $\mathbf{q}^{(\pm)}$ evolve similarly to the coordinates.
In particular, there is a negligible
relative splitting in the $y$ components of momenta, but the $x$ and $z$ components of $\mathbf{q}^{(\pm)}$ oscillate
with time. Unlike the coordinates, however, the average splitting of the $x$ and $z$ components of momenta
does not increase with time. In this connection, we should remark that the wave packet energies never exceed the
threshold value (\ref{model-epsk0-pm}) and, thus, the numerical analysis remains within the range of applicability of
the low-energy theory.

\begin{figure}[t]
\begin{center}
\includegraphics[width=0.59\textwidth]{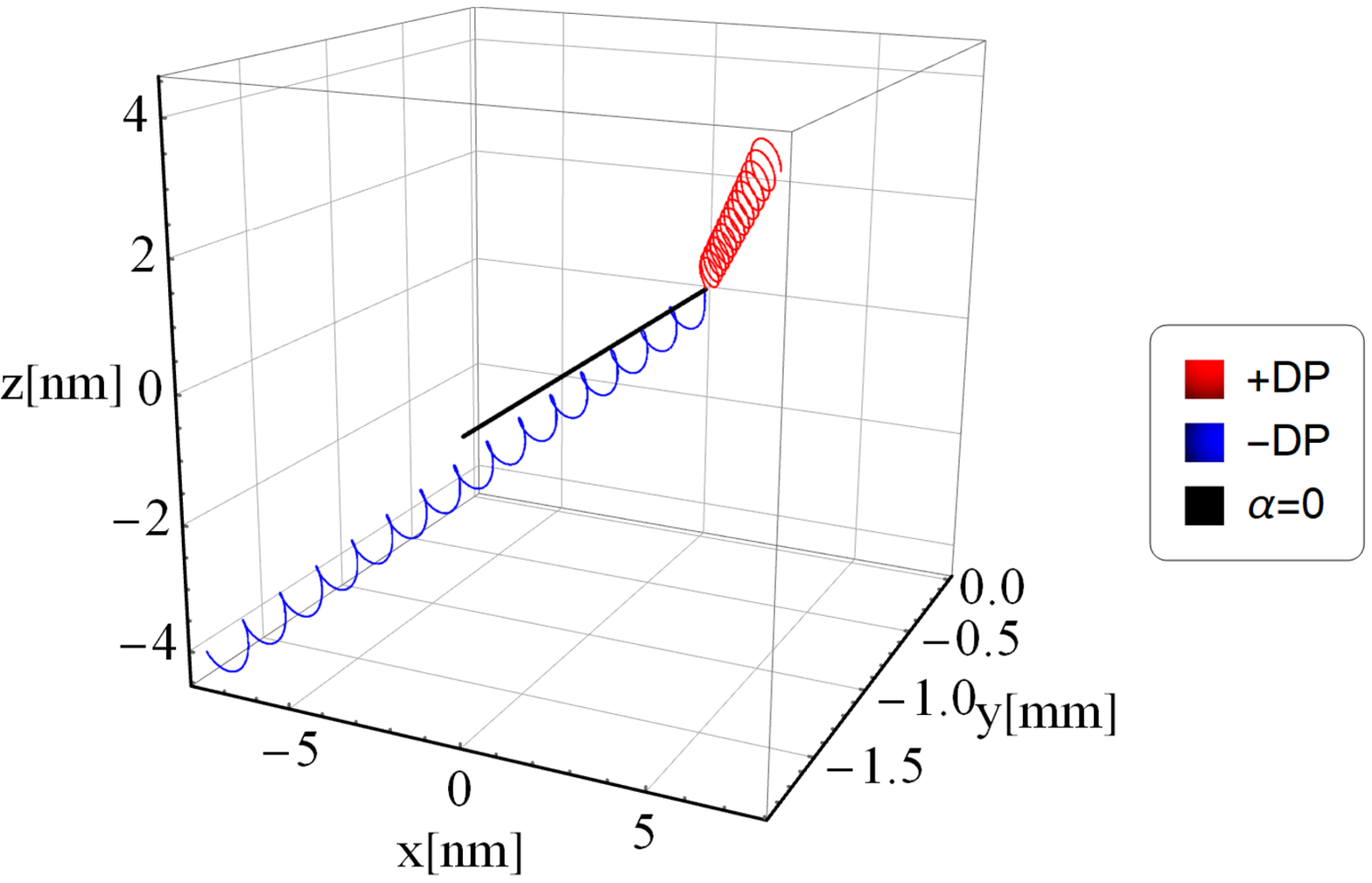}\hfill
\includegraphics[width=0.39\textwidth]{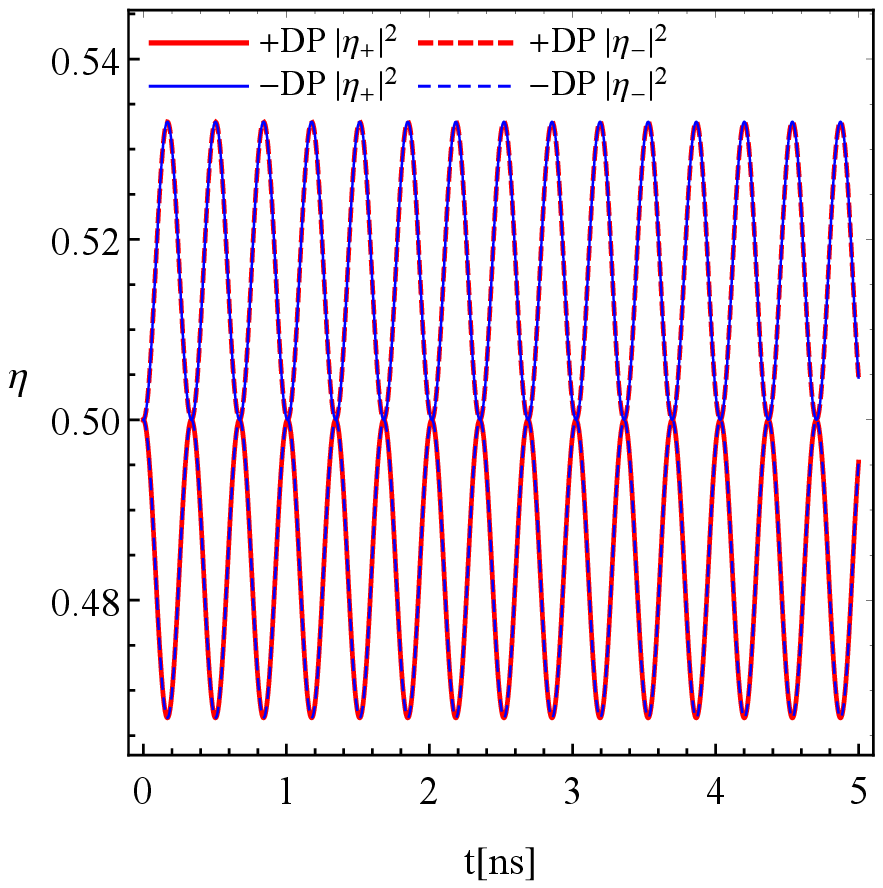}
\caption{Left panel: The trajectories of the wave packets from different Dirac points for $t\leq t_{\rm max}=5~\mbox{ns}$.
The red and blue lines represent the wave packets described by Hamiltonians (\ref{model-Hamiltonian-canonical-plus})
and (\ref{model-Hamiltonian-canonical-minus}) at $\alpha=0.5\alpha^{*}$, respectively. The black line corresponds to
the case $\alpha=0$, where the wave packets do not split.
Right panel: The time dependence of the probabilities $|\eta_{\pm}|^2$ to find the wave packets in
certain $\Gamma_5$-chirality states. The red and blue lines correspond to the wave packets described by
Hamiltonians (\ref{model-Hamiltonian-canonical-plus}) and (\ref{model-Hamiltonian-canonical-minus}).
The solid and dashed lines describe $|\eta_{+}|^2$ and $|\eta_{-}|^2$, respectively.
We used $\mathbf{E}=E\hat{\mathbf{y}}$ and $\mathbf{B}=B\hat{\mathbf{y}}$, where
$E=200~\mbox{V/m}$ and $B=10~\mbox{G}$.}
\label{fig:trajectories-pm-DP-exact-tau-2-Ey-By-r-3D}
\end{center}
\end{figure}

\begin{figure}[t]
\begin{center}
\hspace{-0.32\textwidth}(a)\hspace{0.32\textwidth}(b)\hspace{0.32\textwidth}(c)\\[0pt]
\includegraphics[width=0.32\textwidth]{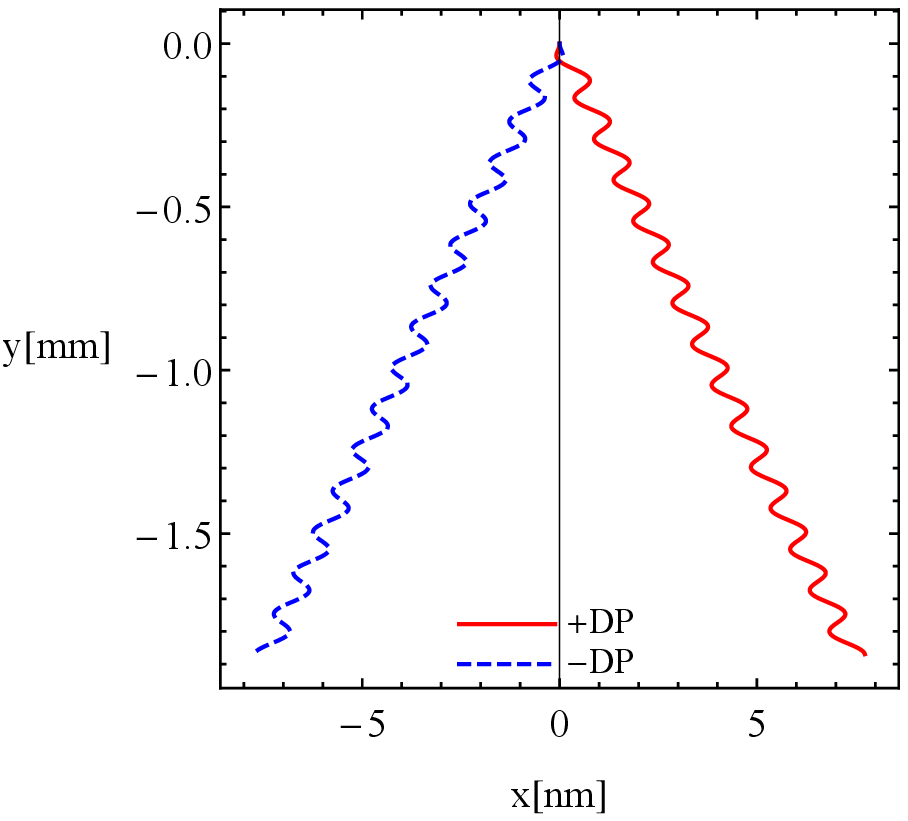}\hfill
\includegraphics[width=0.32\textwidth]{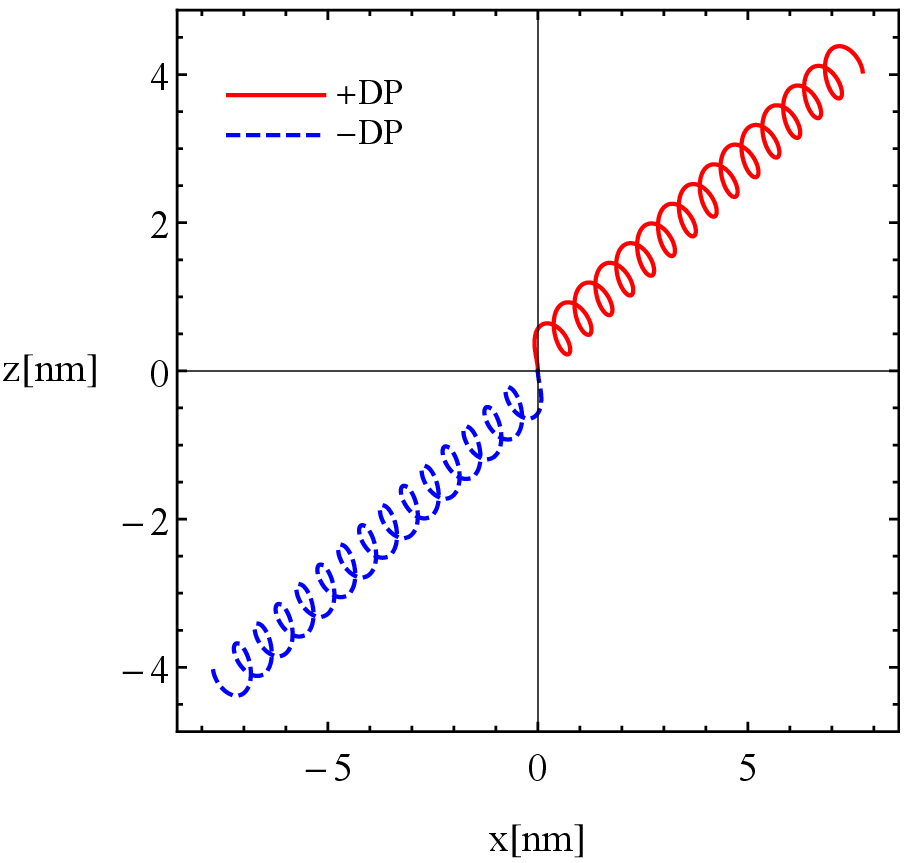}\hfill
\includegraphics[width=0.32\textwidth]{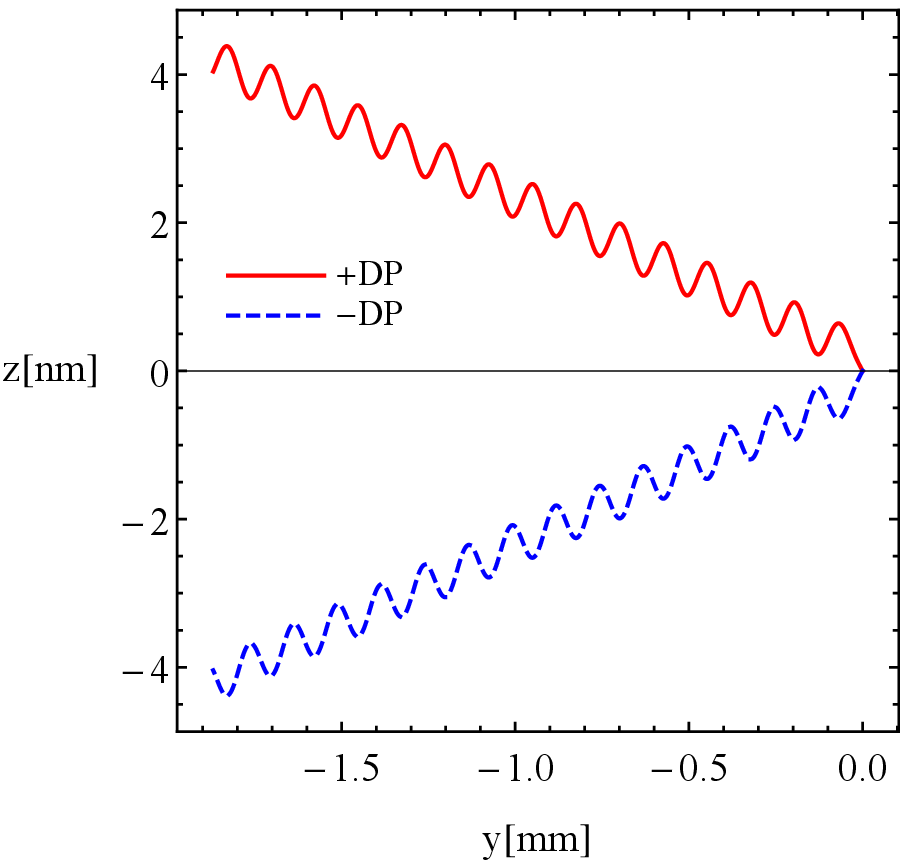}
\caption{The projections of the wave packet trajectories onto the following planes: $x$-$y$ (panel a),
$x$-$z$ (panel b), and $y$-$z$ (panel c). The red and blue lines correspond to the wave packets described
by Hamiltonians (\ref{model-Hamiltonian-canonical-plus}) and (\ref{model-Hamiltonian-canonical-minus}), respectively.
We used $t\leq t_{\rm max}=5~\mbox{ns}$, $\alpha=0.5\alpha^{*}$, $\mathbf{E}=E\hat{\mathbf{y}}$, and
$\mathbf{B}=B\hat{\mathbf{y}}$, where $E=200~\mbox{V/m}$ and $B=10~\mbox{G}$.}
\label{fig:trajectories-pm-DP-exact-tau-2-Ey-By-r-2D}
\end{center}
\end{figure}

The trajectories of the wave packets and the probabilities $|\eta_{\pm}|^2$ for the wave packets from different
valleys to be in certain $\Gamma_5$-chirality states  are presented in the left and right panels of
Fig.~\ref{fig:trajectories-pm-DP-exact-tau-2-Ey-By-r-3D}, respectively. The projections of trajectories onto the
$x$-$y$, $x$-$z$, and $y$-$z$ planes are shown in the three panels of Fig.~\ref{fig:trajectories-pm-DP-exact-tau-2-Ey-By-r-2D}.
The timescale is set to $t\leq t_{\rm max}=5~\mbox{ns}$.
In agreement with the results in Fig.~\ref{fig:trajectories-pm-DP-exact-tau-2-Ey-By-r}, the trajectories
for the wave packets from different Dirac points are clearly separated. The origin of the splitting of the wave packets
from different valleys in the $x$ and $z$ directions, seen Figs.~\ref{fig:trajectories-pm-DP-exact-tau-2-Ey-By-r-3D}
and \ref{fig:trajectories-pm-DP-exact-tau-2-Ey-By-r-2D}, can be traced back to the nontrivial structure of the low-energy
Hamiltonians (\ref{model-Hamiltonian-canonical-plus}) and (\ref{model-Hamiltonian-canonical-minus}). It is remarkable
that the magnitude of the average splitting linearly increases with time. By making use of this fact, we estimate
that the spatial separation can reach a few micrometers for a centimeter-size crystal, provided the latter is sufficiently
clean and the quasiparticle mean free path is sufficiently long. Such a splitting could provide an observational signature
for the nontrivial wave packets dynamics in the Dirac semimetals $\mathrm{A_3Bi}$ (A=Na,K,Rb). One should note,
however, that the splitting is largely washed away in strong magnetic fields (e.g., $B=100~\mbox{G}$) when the
Lorentz force starts to dominate and causes the trajectories to overlap. [Note that the separation of the wave packets can be resolved experimentally only on the spatial scales larger than wave packet's characteristic sizes, i.e., $\lambda \gtrsim 2\pi/k$.]

The small spiral-like features on top of the linear separation in Fig.~\ref{fig:trajectories-pm-DP-exact-tau-2-Ey-By-r-2D} (see also the left panel in Fig.~\ref{fig:trajectories-pm-DP-exact-tau-2-Ey-By-r-3D})
can be traced back to the oscillations of the partial weights $\eta_{\pm}$. This is also confirmed by the results for
$|\eta_{\pm}|^2$ in the right panel of Fig.~\ref{fig:trajectories-pm-DP-exact-tau-2-Ey-By-r-3D}, which demonstrate
that the propagation of the wave packets is accompanied by a weakly oscillating splitting of the $\Gamma_5$-chirality.
From a physics viewpoint, these oscillations are related to the precession of the magnetic moment of the wave packet.
They are determined by the non-Abelian nature of the Berry curvature and the nontrivial structure of the magnetic
moment. From an observational viewpoint, however, these features could be very difficult to detect. Indeed, while the
oscillations could be made larger by increasing the magnetic field, the valley separation becomes weak in such
a regime.

Before proceeding to the case of the perpendicular electric and magnetic fields, let us provide some
underlying reasons for the spatial valley separation of wave packets. Because of a rather complicated
structure of Eqs.~(\ref{WPE-r-eq-exp}), (\ref{WPE-q-eq-exp}), and (\ref{WPE-eta-eq-exp}), we
present only a rough qualitative description. To start with, we assume that the changes of the
partial weights are negligible, i.e., $\eta_{\pm}(t)=const$. In such a case, the spiral-like motion on
top of the mostly linear splitting disappears. Then, it is easy to check that the remaining spatial
separation is driven primarily by the velocity term in Eq.~(\ref{WPE-r-eq-exp}), i.e., the first term
on the right-hand side. In fact, the separation in both $x$ and $z$ directions is related to the same $z$
component of velocity $\mathbf{v}$. While the effect of $v_z$ on the motion in the $z$ direction
is obvious, the splitting in the $x$ direction is achieved indirectly. In particular, the $x$ component of the
velocity is mainly determined by the corresponding component of the momentum, which, in turn, is
generated by the Lorentz force $e v_zB_y/c$ in Eq.~(\ref{WPE-q-eq-exp}), i.e., the second term in expression (\ref{force}). Obviously, such a splitting is induced only when $\mathbf{B}\neq \mathbf{0}$.
However, the presence of nonzero $\alpha$ in the energy dispersion relation (\ref{energy-dispersion-lin})
plays a key role as well: it gives a nonzero $v_z$ everywhere away from the Dirac points and makes
an efficient spatial separation of the wave packets possible. Thus, the valley splitting of wave packets
is in large part connected with the special form of the momentum-dependent chirality-mixing
term $\Delta(\mathbf{q})$ in the low-energy Hamiltonian.

\subsection{Perpendicular electric and magnetic fields}
\label{sec:trajectories-pm-DP-exact-tau-2-Ey-Bz}

In this subsection, we consider the motion of the wave packets in perpendicular electric and magnetic fields.
We use the same magnitudes of the electric and magnetic fields as in the previous subsection, but the
magnetic field is now in the $z$ direction, i.e., $\mathbf{B}=B\hat{\mathbf{z}}$.
Again, $\alpha=0.5\alpha^{*}$, which is sufficiently small to ensure that the relative contribution of the off-diagonal terms, quantified by $\Delta(\mathbf{q})/(v_Fq)$, would remain small for the timescales used in our numerical calculations.

Let us note also that, because of the off-diagonal gap term
$\propto\alpha \sqrt{m} k_{\pm}^2$, the dynamics for the two possible orientations of the magnetic field,
i.e., $\mathbf{B}=B\hat{\mathbf{z}}$ and
$\mathbf{B}=B\hat{\mathbf{x}}$, are not equivalent. In fact, for sufficiently large
timescales, the semiclassical approximation fails in the latter case. Therefore, in this study, we will not
discuss it.

\begin{figure}[t]
\begin{center}
\includegraphics[width=0.45\textwidth]{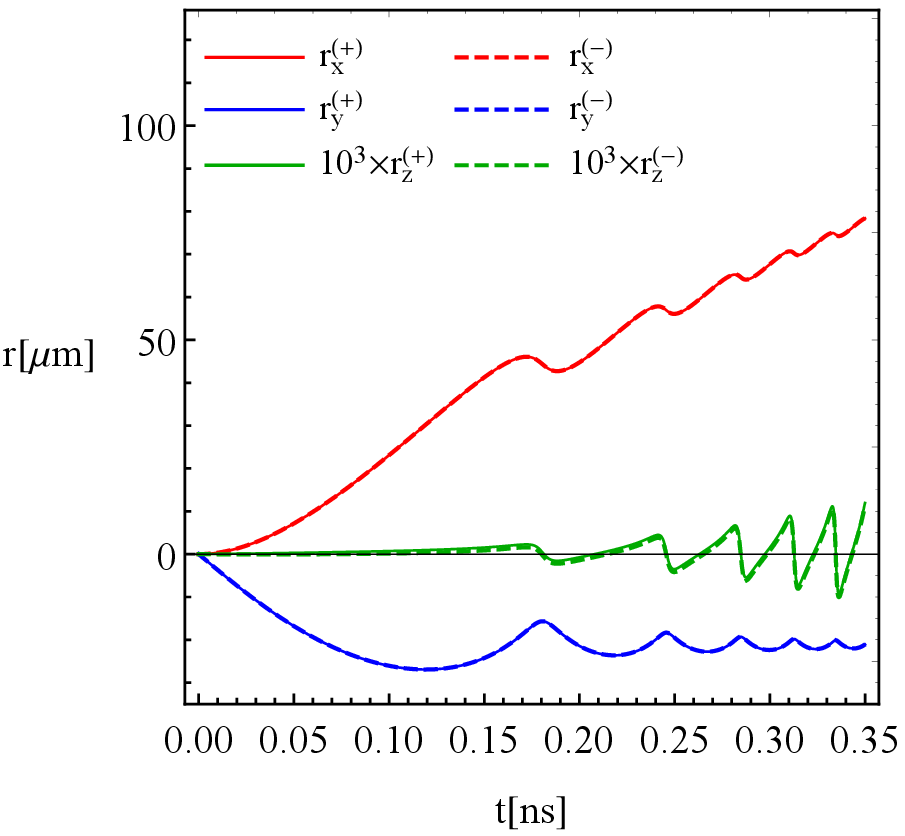}\hfill
\includegraphics[width=0.45\textwidth]{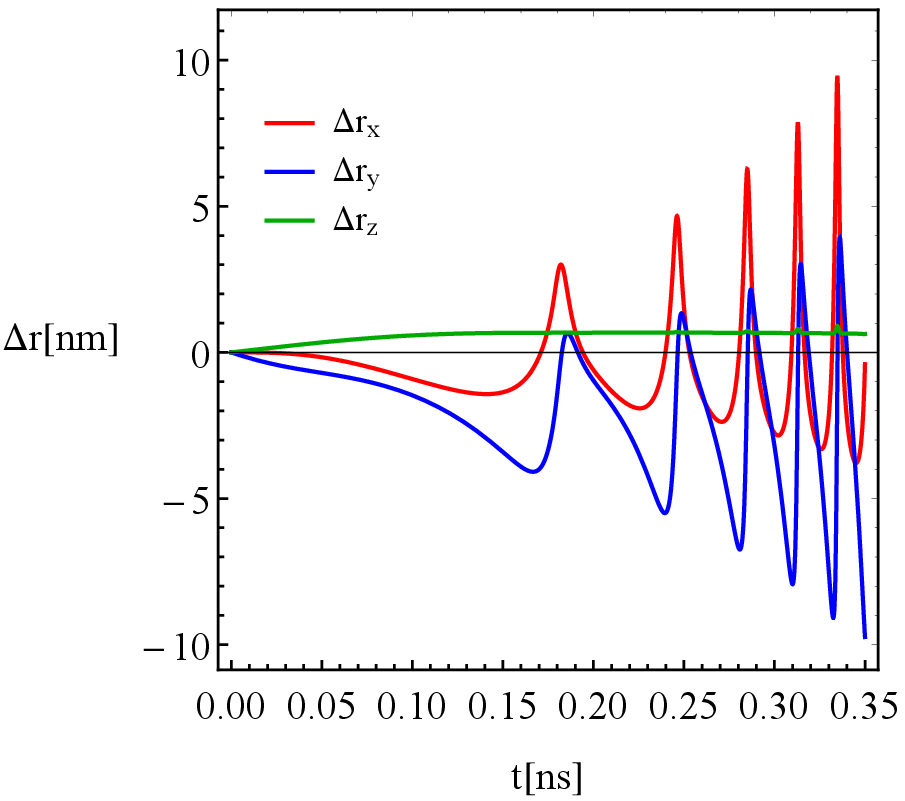}\\
\caption{The positions $\mathbf{r}^{(\pm)}$ of the wave packets as a function of time $t$ (left panel) and the splitting
$\Delta \mathbf{r}=\mathbf{r}^{(+)}-\mathbf{r}^{(-)}$  of the coordinates of the wave packets from different Dirac points
(right panel). The red, blue, and green lines correspond to the $x$, $y$, and $z$ components, respectively.
The solid and dashed lines represent the results for the wave packets described by Hamiltonians
(\ref{model-Hamiltonian-canonical-plus}) and (\ref{model-Hamiltonian-canonical-minus}), respectively.
We used $\alpha=0.5\alpha^{*}$, $\mathbf{E}=E\hat{\mathbf{y}}$ and $\mathbf{B}=B\hat{\mathbf{z}}$, where
$E=200~\mbox{V/m}$ and $B=10~\mbox{G}$.}
\label{fig:trajectories-pm-DP-exact-tau-2-Ey-Bz-r}
\end{center}
\end{figure}

The evolution of the positions $\mathbf{r}^{(\pm)}$ for the wave packets from different Dirac points (valleys)
is shown in Fig.~\ref{fig:trajectories-pm-DP-exact-tau-2-Ey-Bz-r}. As expected in the perpendicular electric
and magnetic fields, the coordinates of the wave packets oscillate in the plane normal to $\mathbf{B}$ (i.e.,
the $x$ and $y$ coordinates), albeit have a nonharmonic pattern. We also found that the non-Abelian
terms lead to the oscillation-like motion of the wave packets along the $z$ axis, as well as to a small splitting
of the trajectories of the wave packets from different Dirac points (see the right panel in
Fig.~\ref{fig:trajectories-pm-DP-exact-tau-2-Ey-Bz-r}). We checked that the wave packet momenta
oscillate too and split slightly when $\alpha\neq0$. Remarkably, however, the $z$ components of momenta
vanish. We conclude, therefore, that the slow motion of the wave packets in the $z$ direction is caused exclusively
by the non-Abelian effects. In all cases presented, we verified that the energies of the wave packets remain
sufficiently small to justify the use of the linearized model.

\begin{figure}[t]
\begin{center}
\includegraphics[width=0.59\textwidth]{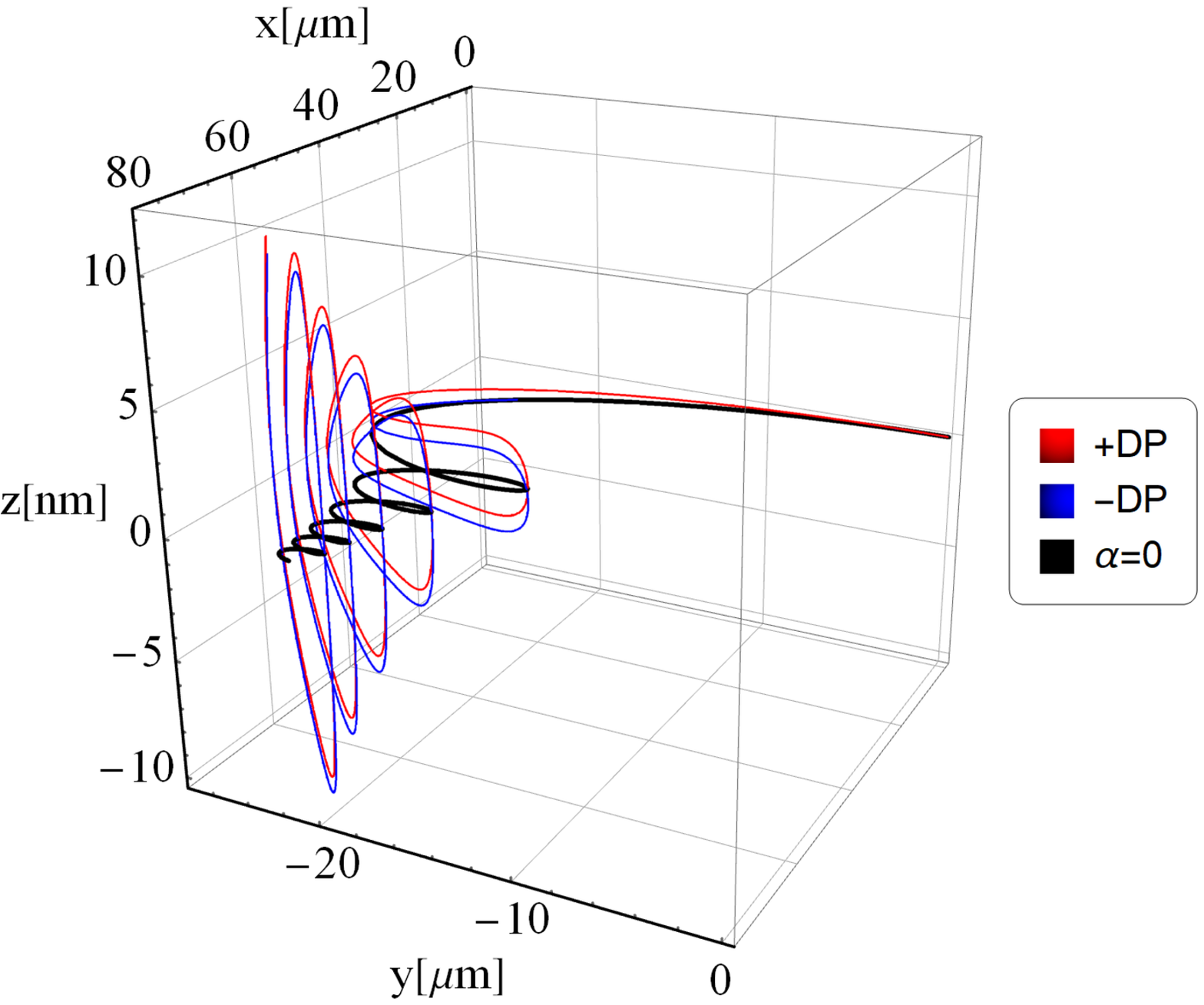}\hfill
\includegraphics[width=0.39\textwidth]{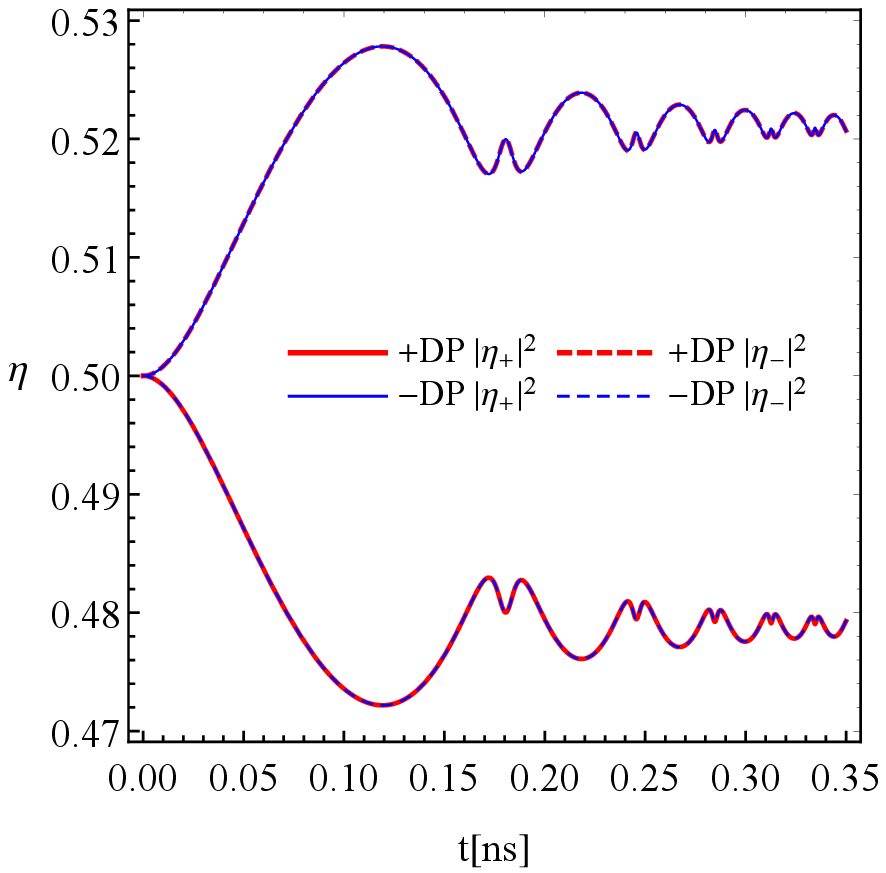}
\caption{Left panel: The trajectories of the wave packets corresponding to different Dirac points or valleys for $t\leq t_{\rm max}=0.35~\mbox{ns}$.
The red and blue lines represent the wave packets described by Hamiltonians (\ref{model-Hamiltonian-canonical-plus}) and (\ref{model-Hamiltonian-canonical-minus}) at $\alpha=0.5\alpha^{*}$, respectively. The black line corresponds to the case $\alpha=0$, where the wave packets are not split.
Right panel: The probabilities $|\eta_{\pm}|^2$ to find the wave packets in certain $\Gamma_5$ states for $\alpha=0.5\alpha^{*}$.
The red and blue lines correspond to the wave packets described by Hamiltonians (\ref{model-Hamiltonian-canonical-plus}) and (\ref{model-Hamiltonian-canonical-minus}).
The solid and dashed lines describe $|\eta_{+}|^2$ and $|\eta_{-}|^2$, respectively.
We used $\mathbf{E}=E\hat{\mathbf{y}}$ and $\mathbf{B}=B\hat{\mathbf{z}}$, where
$E=200~\mbox{V/m}$ and $B=10~\mbox{G}$.}
\label{fig:trajectories-pm-DP-exact-tau-2-Ey-Bz-r-3D}
\end{center}
\end{figure}

\begin{figure}[ht]
\begin{center}
\hspace{-0.32\textwidth}(a)\hspace{0.32\textwidth}(b)\hspace{0.32\textwidth}(c)\\[0pt]
\includegraphics[width=0.32\textwidth]{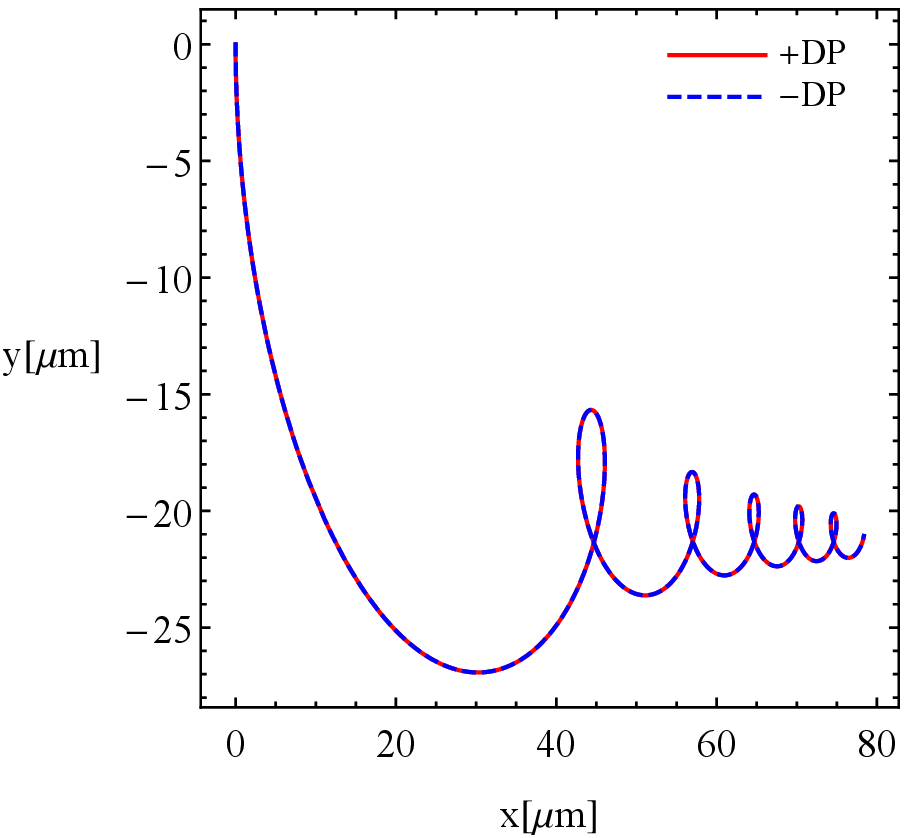}\hfill
\includegraphics[width=0.32\textwidth]{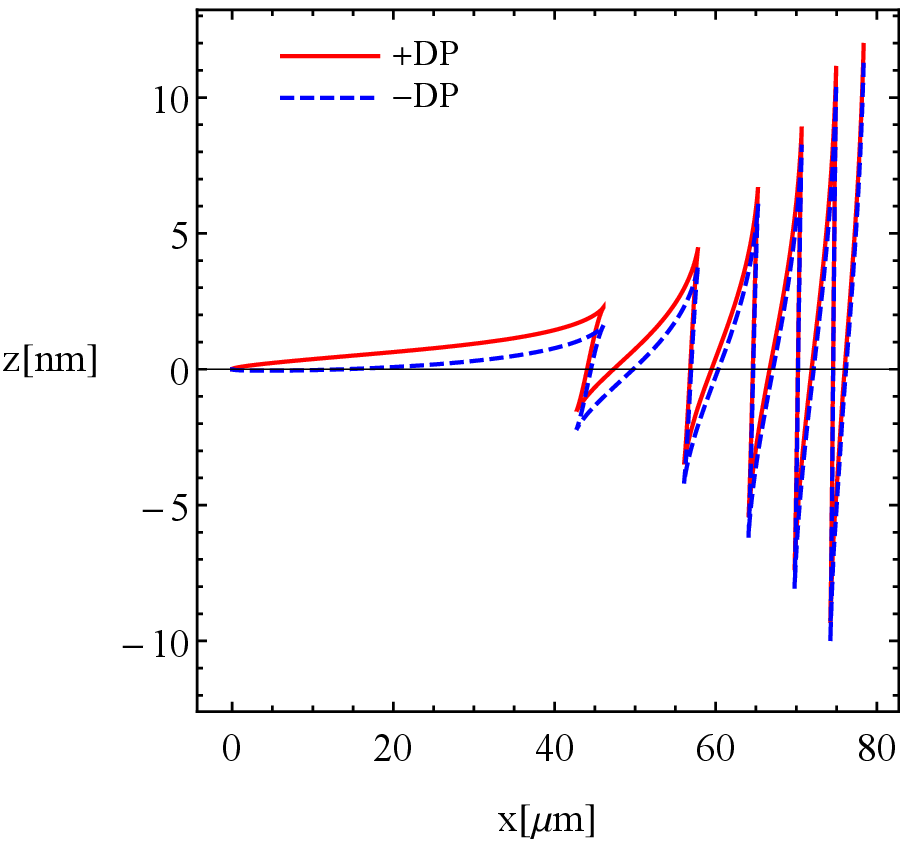}\hfill
\includegraphics[width=0.32\textwidth]{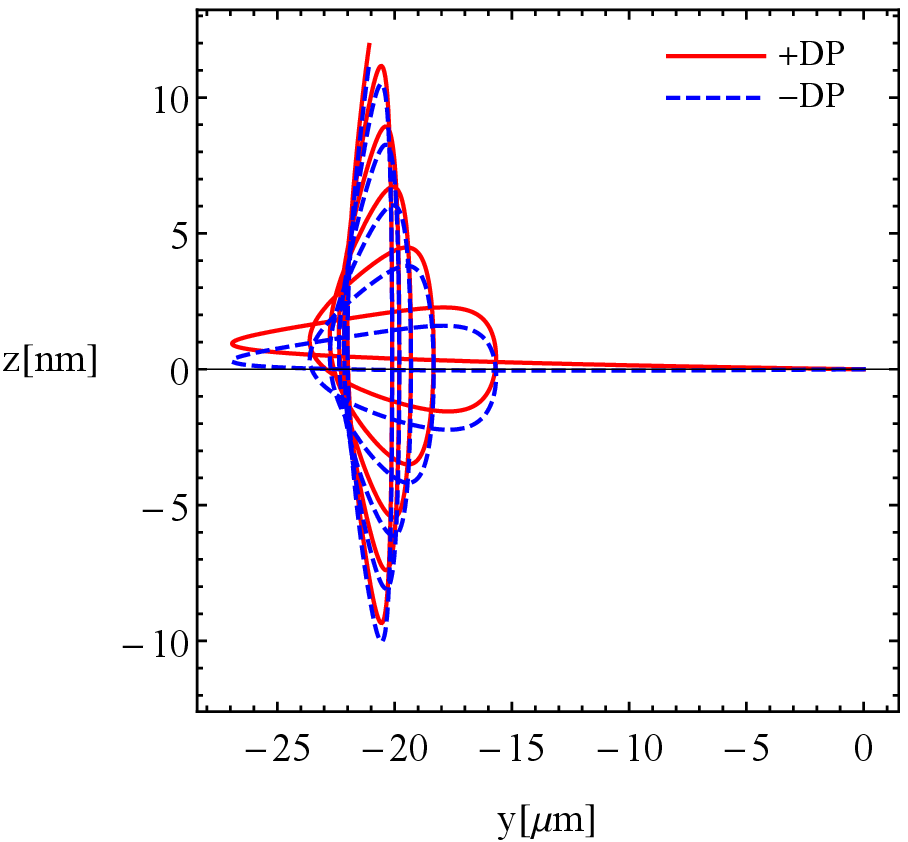}
\caption{The projections of the wave packets trajectories onto the following planes:
$x$-$y$ (panel a), $x$-$z$ (panel b), and $y$-$z$ (panel c). The red solid and blue
dashed lines correspond to the wave packets described by Hamiltonians
(\ref{model-Hamiltonian-canonical-plus}) and (\ref{model-Hamiltonian-canonical-minus}), respectively.
We used $t\leq t_{\rm max}=0.35~\mbox{ns}$, $\alpha=0.5\alpha^{*}$,
$\mathbf{E}=E\hat{\mathbf{y}}$, and $\mathbf{B}=B\hat{\mathbf{z}}$, where
$E=200~\mbox{V/m}$ and $B=10~\mbox{G}$.}
\label{fig:trajectories-pm-DP-exact-tau-2-Ey-Bz-r-2D}
\end{center}
\end{figure}

We present the trajectories of the wave packets and the probabilities $|\eta_{\pm}|^2$ to find the
wave packets from different valleys in certain chiral states in the left and right panels of
Fig.~\ref{fig:trajectories-pm-DP-exact-tau-2-Ey-Bz-r-3D}, respectively. The projections of the
wave packet trajectories onto the $x$-$y$, $x$-$z$, and $y$-$z$ planes are shown in the three
panels of Fig.~\ref{fig:trajectories-pm-DP-exact-tau-2-Ey-Bz-r-2D}. The timescale is set to
$t\leq t_{\rm max}=0.35~\mbox{ns}$.

As is clear from the results in the left panel of Fig.~\ref{fig:trajectories-pm-DP-exact-tau-2-Ey-Bz-r-3D},
the non-Abelian corrections lead to rapid oscillations of the wave packets in the $z$ direction. Note
that while the amplitude of oscillations increases, their period decreases with time. Such a behavior
suggests that the quasiclassical approximation gradually breaks down. The spatial oscillations of the
wave packets can be traced back to an oscillatory time dependence of the partial weights and disappear if
one enforces constant weights $\eta_{\pm}(t)$.
In general, the trajectories of the wave packets from
different Dirac points are asymmetric with respect to the
$x$-$y$ plane. On the other hand, by taking into account their substantial overlap (see the left panel of Fig.~\ref{fig:trajectories-pm-DP-exact-tau-2-Ey-Bz-r-3D}),
we think that the valley separation cannot be easily achieved in this case. We checked, however,
that trajectories change qualitatively at sufficiently large magnetic fields and the valley separation
in the $z$ direction becomes possible at least in principle, although its magnitude is estimated to be
rather small.

It is interesting to point that, according to the right panel in Fig.~\ref{fig:trajectories-pm-DP-exact-tau-2-Ey-Bz-r-3D},
the wave packets from different Dirac points develop nonzero and opposite in sign $\Gamma_5$-chirality
polarizations. Such polarizations have an interesting oscillatory pattern with the absolute values of the
partial weights reaching almost constant values at sufficiently large timescales. In summary, while the
valley separation is weak, the deviation of the wave packets from the $x$-$y$ plane clearly provides an
evidence for the non-Abelian effects.

\subsection{Motion of wave packets for chirally polarized initial states}
\label{sec:trajectories-pm-DP-exact-tau-2-polarization}

In this subsection, for completeness, we investigate the motion of wave packets when the initial states
are chirally polarized. We limit ourselves to the two limiting configurations given by
Eqs.~(\ref{trajectories-pm-DP-exact-tau-2-polarization-eta-1}) and (\ref{trajectories-pm-DP-exact-tau-2-polarization-eta-2}).

\begin{figure}[ht]
\begin{center}
\includegraphics[width=0.75\textwidth]{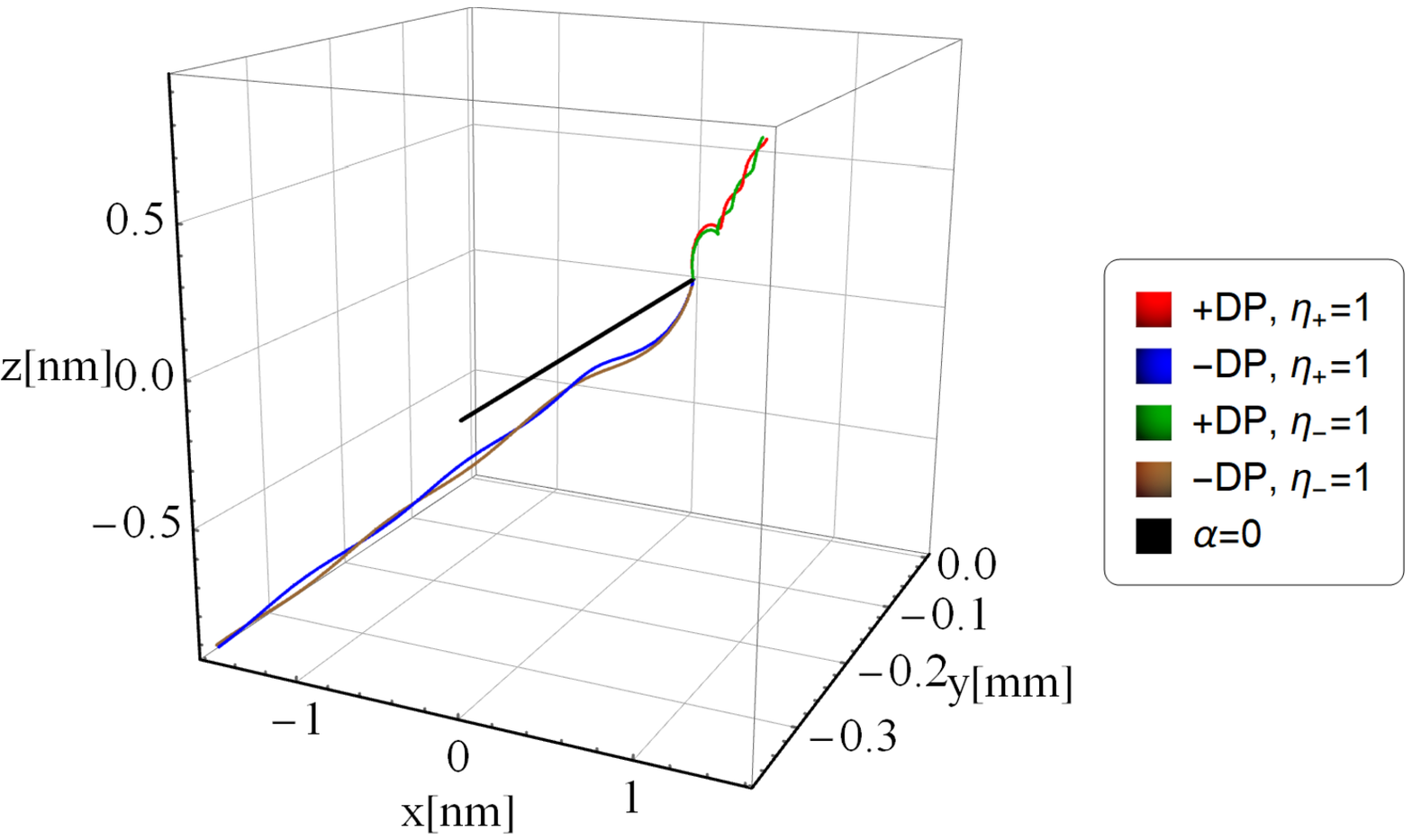}
\caption{The trajectories of the wave packets in parallel electric and magnetic fields.
The red and blue lines correspond to the wave packets for the initial weights
(\ref{trajectories-pm-DP-exact-tau-2-polarization-eta-1}) described by Hamiltonians
(\ref{model-Hamiltonian-canonical-plus}) and (\ref{model-Hamiltonian-canonical-minus}), respectively.
The green and brown lines correspond to the wave packets for the initial weights
(\ref{trajectories-pm-DP-exact-tau-2-polarization-eta-2}) described by Hamiltonians
(\ref{model-Hamiltonian-canonical-plus}) and (\ref{model-Hamiltonian-canonical-minus}), respectively.
We used $t\leq t_{\rm max}=1~\mbox{ns}$, $\alpha=0.5\alpha^{*}$, $\mathbf{E}=E\hat{\mathbf{y}}$ and $\mathbf{B}=B\hat{\mathbf{y}}$,
where $E=200~\mbox{V/m}$ and $B=10~\mbox{G}$.}
\label{fig:trajectories-pm-DP-exact-tau-2-polarization-Ey-By-r-3D}
\end{center}
\end{figure}

\begin{figure}[ht]
\begin{center}
\hspace{-0.32\textwidth}(a)\hspace{0.32\textwidth}(b)\hspace{0.32\textwidth}(c)\\[0pt]
\includegraphics[width=0.32\textwidth]{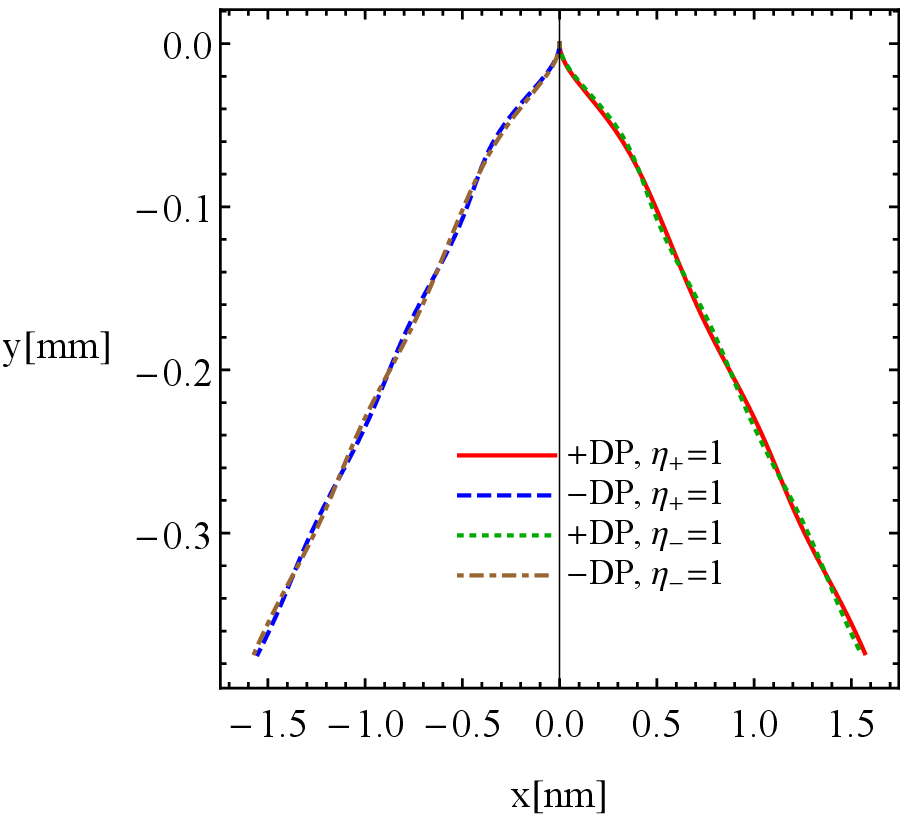}\hfill
\includegraphics[width=0.32\textwidth]{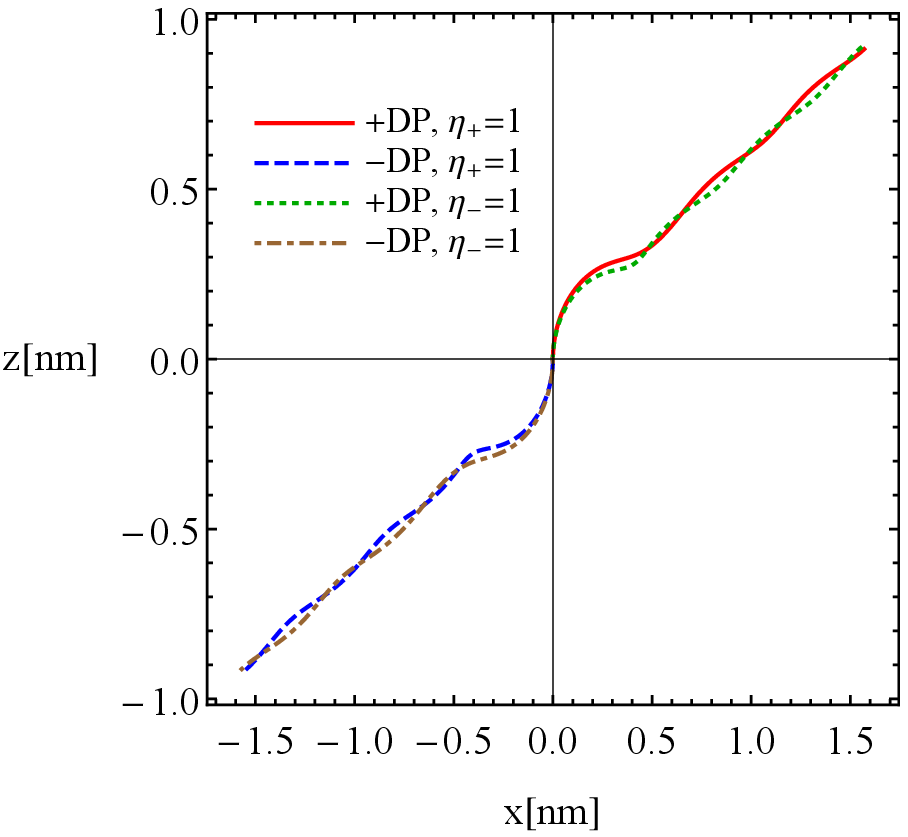}\hfill
\includegraphics[width=0.32\textwidth]{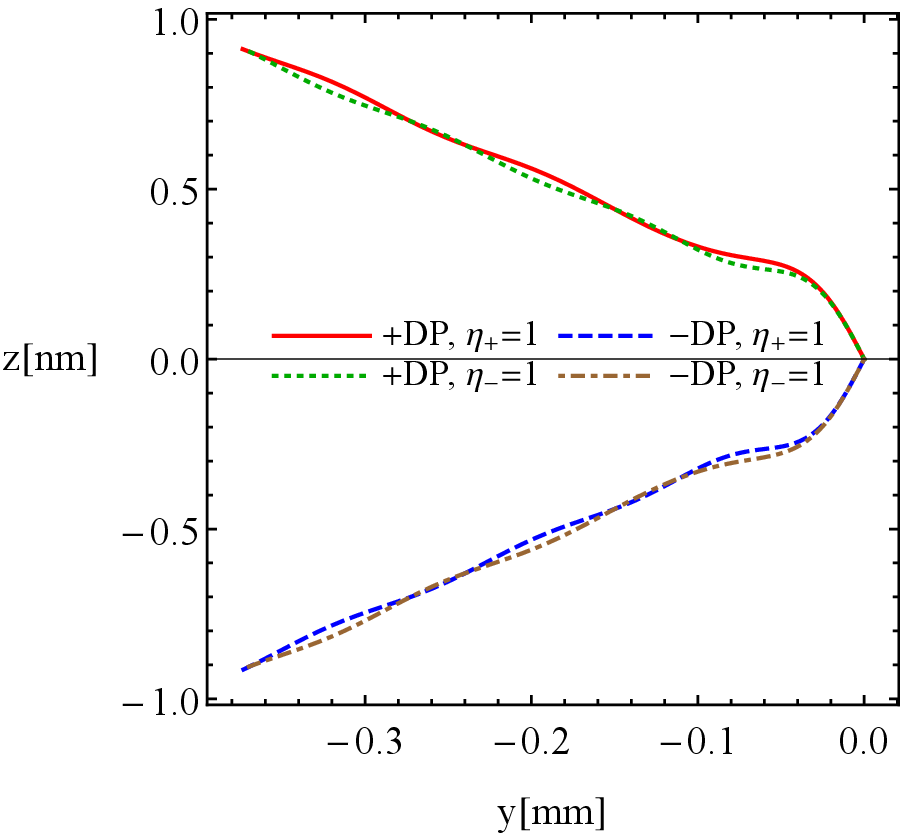}
\caption{The projections of the wave packet trajectories onto the following planes: $x$-$y$ (panel a), $x$-$z$ (panel b), and $y$-$z$ (panel c).
The red solid and blue dashed lines correspond to the wave packets for the initial weights
(\ref{trajectories-pm-DP-exact-tau-2-polarization-eta-1}) described by Hamiltonians (\ref{model-Hamiltonian-canonical-plus}) and
(\ref{model-Hamiltonian-canonical-minus}), respectively.
The green and brown lines correspond to their counterparts with the initial weights (\ref{trajectories-pm-DP-exact-tau-2-polarization-eta-2}).
We used $t\leq t_{\rm max}=1~\mbox{ns}$, $\alpha=0.5\alpha^{*}$, $\mathbf{E}=E\hat{\mathbf{y}}$ and $\mathbf{B}=B\hat{\mathbf{y}}$,
where $E=200~\mbox{V/m}$ and $B=10~\mbox{G}$.}
\label{fig:trajectories-pm-DP-exact-tau-2-polarization-Ey-By-r-2D}
\end{center}
\end{figure}

For the sake of brevity, we investigate only the most interesting case of parallel electric and
magnetic fields. The corresponding results are shown in Fig.~\ref{fig:trajectories-pm-DP-exact-tau-2-polarization-Ey-By-r-3D}
with the projections onto the $x$-$y$, $x$-$z$, and $y$-$z$ planes presented in the three panels of
Fig.~\ref{fig:trajectories-pm-DP-exact-tau-2-polarization-Ey-By-r-2D}. The timescale is limited to
$t\leq t_{\rm max}=1~\mbox{ns}$. While the probabilities to find wave packets in states with fixed
chirality are not shown, we checked that they weakly oscillate around their initial values. Just like
in the case of the non-chiral wave packets discussed in Sec.~\ref{sec:trajectories-pm-DP-exact-tau-2-Ey-By},
the physical origin of this subdominant oscillating motion can be traced to the precession of the magnetic moment.
As one can easily see from Figs.~\ref{fig:trajectories-pm-DP-exact-tau-2-polarization-Ey-By-r-3D} and
\ref{fig:trajectories-pm-DP-exact-tau-2-polarization-Ey-By-r-2D}, the trajectories of the wave packets
corresponding to different Dirac points but with the same initial $\Gamma_5$ weights are completely
split and the amplitude of the splitting increases with time. On the other hand, the wave packets with
different initial $\Gamma_5$ weights are weakly separated. Therefore, in the case of the nonequal
initial weights~(\ref{trajectories-pm-DP-exact-tau-2-polarization-eta-1}) and
(\ref{trajectories-pm-DP-exact-tau-2-polarization-eta-2}), there is a sufficiently weak splitting of the chiral
wave packets on top of the relatively large valley splitting.

\section{Summary and discussions}
\label{sec:Summary}

In this paper, we investigated the dynamics of the electron wave packets in the Dirac semimetals $\mathrm{A_3Bi}$ (A=Na,K,Rb).
We showed that due to the hidden $Z_2$ Weyl nature of these materials \cite{Gorbar:2014sja}, the semiclassical motion of
the wave packets is qualitatively affected by the non-Abelian contributions when external electric $\mathbf{E}$ and magnetic
$\mathbf{B}$ fields are applied to the system. These contributions arise due to the degeneracy of the electron states and the
chirality-mixing from $\Delta(\mathbf{k})$ in the effective Hamiltonian. Unlike the usual mass (gap) term in the Dirac Hamiltonian,
$\Delta(\mathbf{k})$ is momentum-dependent and vanishes at the Dirac points. As a result, the gapless energy spectrum is
preserved and the chirality remains well defined in the close vicinity of the Dirac points. The doubly degenerate states
near each Dirac point can be classified with respect to the $\Gamma_5$ transformation. [Since at small $\Delta(\mathbf{k})$
the latter is approximately the same as the chiral transformation, we use the term chirality to classify the corresponding
states.]

It is found that when $\mathbf{E}\parallel\mathbf{B}$ and the magnetic field is sufficiently small, the trajectories of
wave packets from different valleys (or, equivalently, Dirac points) are spatially split in the plane perpendicular to the fields.
What is more important, the magnitude of the valley separation grows linearly with time. One might speculate,
therefore, that a substantial splitting could be achieved in macroscopic systems when the quasiparticle mean free
path is sufficiently large. (Note that the propagation of wave packets depends on the relative phase of the weights
whose values, however, would be difficult to control in experiments.) Interestingly, the non-Abelian corrections
allow for a spiral-like motion of the wave packets on top of the almost linear separation. As is clear, the physical
origin of such spiraling is connected with the precession of the magnetic moment. The same effect allows also
for a small oscillating chirality polarization of the wave packets. While the amplitude of the spirals is estimated
to be relatively small, the linear splitting of the trajectories due to the momentum-dependent chirality-mixing term
could reach micrometers for centimeter-size crystals. When the wave packets are initially chirality polarized,
there is a weak splitting of the chiral wave packets on top of the well-pronounced valley separation. The latter
has the same origin as for the nonpolarized wave packets. In the case of a strong magnetic field, the Lorentz force
dominates that leads to weakly separated trajectories of the wave packets from different Dirac points.
Therefore, we believe that the setup with the parallel electric and magnetic fields allows for a spatial splitting
of the wave packets that can be, in principle, tested experimentally.

When the electric and sufficiently weak magnetic fields are perpendicular, the valley separation is negligible
for the equal initial weights of the degenerate chirality states. On the other hand, the non-Abelian
corrections lead to a well-pronounced oscillating motion of the wave packets in the direction parallel to the magnetic field.
If detected, such deviations from the usual in-plane motion could provide another signature of the non-Abelian effects.
In addition, there is also a weak chirality polarization of the states from different Dirac points, which, however,
is not easily accessible because the trajectories from different valleys are not well-split.
The situation changes at sufficiently large magnetic fields, when the trajectories from different valleys are separated
along the direction of the magnetic field. However, the separation is nonmonotonic and is estimated to be relatively weak.
Therefore, while the case of the perpendicular electric and magnetic fields contains interesting physics, it might be difficult
to realize experimentally.

It is instructive to compare the obtained results with those in Ref.~\cite{Gorbar:2017dtp}, where the valley and chirality
splitting was shown to be possible by applying a superposition of magnetic and strain-induced pseudomagnetic fields. In the absence
of the chirality-mixing term $\Delta(\mathbf{k})$ and the non-Abelian corrections to the Berry curvature, however, it was
critical to include a pseudomagnetic field. Without the latter, the right- and left-handed beams from different valleys
would overlap and form nonchiral beams that do not correspond to a certain valley. In contrast, as we showed in the
study here, the non-Abelian effects and the gap term can lead to both valley and chirality splittings even in
the absence of a pseudomagnetic field. While the effects are estimated to be rather small, they can be experimentally accessible via certain local probes.

\begin{acknowledgments}
The work of E.V.G. was partially supported by the Program of Fundamental Research of the
Physics and Astronomy Division of the National Academy of Sciences of Ukraine.
The work of V.A.M. and P.O.S. was supported by the Natural Sciences and Engineering Research Council of Canada.
The work of I.A.S. was supported by the U. S. National Science Foundation under Grants No.~PHY-1404232
and No.~PHY-1713950.
\end{acknowledgments}

\appendix

\section{Explicit expressions for the Berry connection, the Berry curvature, and the magnetic moment}
\label{sec:app-exp-expressions-lin-alpha}

In this appendix, we present the explicit expressions for the Berry connection, the Berry curvature, and the magnetic moment defined in the main
text by Eqs.~(\ref{WPE-Berry-connection-def}), (\ref{WPE-Berry-curvature-def}), and (\ref{WPE-magnetic-moment-def}), respectively.
Unfortunately, the general form of the corresponding results is bulky.
Therefore, we consider only the linear in $\alpha$ approximation.

Let us start from the Berry connection $\mathbf{A}_{mn}^{(\pm)}$ given by Eq.~(\ref{WPE-Berry-connection-def}).
Here, the upper index corresponds to the linearized Hamiltonian given either by Eq.~(\ref{model-Hamiltonian-canonical-plus}) or
Eq.~(\ref{model-Hamiltonian-canonical-minus}) in the main text and $n,m=\pm$ correspond to the wave functions defined in the main text by
Eqs.~(\ref{WPE-psi-def-p}) and (\ref{WPE-psi-def-m}), respectively. The explicit expressions for the Berry connection matrix components are
\begin{eqnarray}
\label{exp-expressions-lin-alpha-A++}
\mathbf{A}_{++}^{(\pm)} &=& \frac{\left(k\pm k_z\right)}{2k_{\perp}^2k} \left\{-k_y, k_x, 0\right\},\\
\label{exp-expressions-lin-alpha-A+-}
\mathbf{A}_{+-}^{(\pm)} &=& \frac{i\alpha \left(\sqrt{m} \pm k_z\right) k_{-}}{2v_Fk_{+}k_{\perp}k} \left\{k_xk_z +2ik_y\left(k_z\mp k\right), k_yk_z
-2ik_x\left(k_z\mp k\right), \pm\frac{k_{\perp}^2 k_z}{\left(\sqrt{m} \pm k_z\right)}\right\},\\
\label{exp-expressions-lin-alpha-A-+}
\mathbf{A}_{-+}^{(\pm)} &=& \left(\mathbf{A}_{+-}^{(\pm)}\right)^{\dag},\\
\label{exp-expressions-lin-alpha-A--}
\mathbf{A}_{--}^{(\pm)} &=& \frac{\left(k\mp k_z\right)}{2k_{\perp}^2k} \left\{-k_y, k_x, 0\right\},
\end{eqnarray}
where $k=\sqrt{k_{\perp}^2+k_z^2}$, $k_{\perp}=\sqrt{k_x^2+k_y^2}$, $v_F$ is the Fermi velocity measured in the energy units, $\sqrt{m}$
determines the separation of the
Dirac points in the momentum space, and the parameter $\alpha$ defines the strength of the off-diagonal terms in Hamiltonians (\ref{model-Hamiltonian-canonical-plus}) and (\ref{model-Hamiltonian-canonical-minus}) in the main text.
It is worth noting that the $z$ component of the Berry connection matrix $\mathbf{A}_{nm}^{(\pm)}$ vanishes when the dependence on $k_z$ is
ignored in $\Delta(\mathbf{k})$.

Further, we present the results for the Berry curvature matrix $\bm{\Omega}_{mn}^{(\pm)}$ given by Eq.~(\ref{WPE-Berry-curvature-def}) in the main text. Its components are
\begin{eqnarray}
\label{exp-expressions-lin-alpha-Omega++}
\bm{\Omega}_{++}^{(\pm)} &=& \mp\frac{\mathbf{k}}{2\hbar k^3},\\
\label{exp-expressions-lin-alpha-Omega+-}
\bm{\Omega}_{+-}^{(\pm)} &=& \frac{\alpha k_{-}}{2v_F\hbar k_{\perp}k_{+} k^3} \Bigg\{-\sqrt{m} k_{\perp}^2 (2k_x+i k_y)\mp 2k_{\perp}^2 k_{+} k_z +k_x k (2k_{\perp}^2+k_z^2), \nonumber\\
&-&\sqrt{m} k_{\perp}^2 (2k_y-i k_x)\pm 2 i k_{\perp}^2 k_{+} k_z +k_y k (2k_{\perp}^2+k_z^2), \nonumber\\
&\mp& (\sqrt{m}\pm k_z)\frac{2k_x^4+2k_y^4+k_z^4+k_y^2k_z(3k_z\pm 2k)+k_x^2 \left[4k_y^2 +k_z(3k_z\pm 2k)\right]}{k}\Bigg\},\\
\label{exp-expressions-lin-alpha-Omega-+}
\bm{\Omega}_{-+}^{(\pm)} &=& \left(\bm{\Omega}_{+-}^{(\pm)}\right)^{\dag},\\
\label{exp-expressions-lin-alpha-Omega--}
\bm{\Omega}_{--}^{(\pm)} &=& \pm\frac{\mathbf{k}}{2\hbar k^3},
\end{eqnarray}
where $\hbar$ is the Planck constant.
Finally, the components of the magnetic moment matrix $\mathbf{M}_{mn}^{(\pm)}$ given by Eq.~(\ref{WPE-magnetic-moment-def}) in the main text read as
\begin{eqnarray}
\label{exp-expressions-lin-alpha-M++}
\mathbf{M}_{++}^{(\pm)} &=& \mp  \frac{ev_F\mathbf{k}}{2\hbar c k^2} = \frac{ev_F k}{c}\bm{\Omega}_{++}^{\pm} ,\\
\label{exp-expressions-lin-alpha-M+-}
\mathbf{M}_{+-}^{(\pm)} &=& \frac{e \alpha k_{-} k_{\perp}}{2\hbar c k_{+} k^2} \Big\{ k_x\left[k-2(\sqrt{m}\pm k_z)\right] -ik_y (\sqrt{m} \pm 2k_z), k_y\left[k-2(\sqrt{m}\pm k_z)\right] +ik_x (\sqrt{m} \pm 2k_z), \nonumber\\
&\mp& (\sqrt{m}\pm k_z) (k\pm 2k_z)
\Big\},\\
\label{exp-expressions-lin-alpha-M-+}
\mathbf{M}_{-+}^{(\pm)} &=& \left(\mathbf{M}_{+-}^{(\pm)}\right)^{\dag},\\
\label{exp-expressions-lin-alpha-M--}
\mathbf{M}_{--}^{(\pm)} &=& \pm\frac{ev_F\mathbf{k}}{2\hbar c k^2} = \frac{ev_F k}{c}\bm{\Omega}_{--}^{\pm}.
\end{eqnarray}
Here, $c$ denotes the speed of light and $e$ is the absolute value of the electron charge.


\begin{thebibliography}{100}

	\bibitem{Fang} Z.~Wang, Y.~Sun, X.~Q.~Chen, C.~Franchini, G.~Xu, H.~Weng, X.~Dai, and Z.~Fang,
Phys. Rev. B {\bf 85}, 195320 (2012).

	\bibitem{WangWeng} Z.~Wang, H.~Weng, Q.~Wu, X.~Dai, and Z.~Fang,
    Phys. Rev. B {\bf 88}, 125427 (2013).

	\bibitem{Borisenko} S.~Borisenko, Q.~Gibson, D.~Evtushinsky, V.~Zabolotnyy, B.~Buchner, and R.~J.~Cava,
Phys. Rev. Lett. {\bf 113}, 027603 (2014).

	\bibitem{Neupane} M.~Neupane, S.-Y.~Xu, R.~Sankar, N.~Alidoust, G.~Bian, C.~Liu, I.~Belopolski, T.-R.~Chang,
H.-T.~Jeng, H.~Lin, A.~Bansil, F.~Chou, and M.~Z.~Hasan,
Nature Commun. {\bf 5}, 3786 (2014).

	\bibitem{Liu} Z.~K.~Liu, B.~Zhou, Y.~Zhang, Z.~J.~Wang, H.~M.~Weng, D.~Prabhakaran, S.-K.~Mo, Z.~X.~Shen,
Z.~Fang, X.~Dai, Z.~Hussain, and Y.~L.~Chen,
Science {\bf 343}, 864 (2014).

	\bibitem{Savrasov} X.~Wan, A.~M.~Turner, A.~Vishwanath, and S.~Y.~Savrasov,
Phys. Rev. B {\bf 83}, 205101 (2011).

	\bibitem{Tong} 
C.-L.~Zhang, Z.~Yuan, Q.-D.~Jiang, B.~Tong, C.~Zhang, X.~C.~Xie, and S.~Jia
    Phys. Rev. B {\bf 95}, 085202 (2017). 

	\bibitem{Bian} S.-Y.~Xu, I.~Belopolski, N.~Alidoust, M.~Neupane, C.~Zhang, R.~Sankar, S.-M.~Huang, C.-C.~Lee, G.~Chang, B.~Wang, G.~Bian, H.~Zheng, D.~S.~Sanchez, F.~Chou, H.~Lin, S.~Jia, and M.~Z.~Hasan,
  Science {\bf 349}, 613 (2015).   

	\bibitem{Qian} B.~Q.~Lv, H.~M.~Weng, B.~B.~Fu, X.~P.~Wang, H.~Miao, J.~Ma, P.~Richard, X.~C.~Huang, L.~X.~Zhao, G.~F.~Chen, Z.~Fang, X.~Dai, T.~Qian, and H.~Ding,
  Phys. Rev. X {\bf 5}, 031013 (2015). 

	\bibitem{Long} X.~Huang, L.~Zhao, Y.~Long, P.~Wang, D.~Chen, Z.~Yang, H.~Liang, M.~Xue, H.~Weng, Z.~Fang, X.~Dai, and G.~Chen,
  Phys. Rev. X {\bf 5}, 031023 (2015). 

	\bibitem{Xu-Hasan:TaP} S.-Y.~Xu, I.~Belopolski, D.~S.~Sanchez, C.~Zhang, G.~Chang, C.~Guo, G.~Bian, Z.~Yuan, H.~Lu, T.-R.~Chang, P.~P.~Shibayev, M.~L.~Prokopovych, N.~Alidoust, H.~Zheng, C.-C.~Lee, S.-M.~Huang, R.~Sankar, F.~Chou, C.-H.~Hsu, H.-T.~Jeng, \emph{et al.},
Sci. Adv. {\bf 1}, e1501092 (2015).

	\bibitem{Xu-Hasan:NbAs} S.-Y.~Xu, N.~Alidoust, I.~Belopolski, Z.~Yuan, G.~Bian, T.-R.~Chang, H.~Zheng, V.~N.~Strocov, D.~S.~Sanchez, G.~Chang, C.~Zhang, D.~Mou, Y.~Wu, L.~Huang, C.-C.~Lee, S.-M.~Huang, B.~Wang, A.~Bansil, H.-T.~Jeng, T.~Neupert, A.~Kaminski, H.~Lin, S.~Jia, and M.~Z.~Hasan,
Nature Phys. {\bf 11}, 748 (2015).

	\bibitem{Xu-Feng:NbP} D.-F.~Xu, Y.-P.~Du, Z.~Wang, Y.-P.~Li, X.-H.~Niu, Q.~Yao, P.~Dudin, Z.-A.~Xu, X.-G.~Wan, and D.-L.~Feng,
Chin. Phys. Lett. {\bf 32}, 107101 (2015).

	\bibitem{Shekhar-Nayak:2015} C.~Shekhar, A.~K.~Nayak, Y.~Sun, M.~Schmidt, M.~Nicklas, I.~Leermakers, U.~Zeitler, Y.~Skourski, J.~Wosnitza, Z.~Liu, Y.~Chen, W.~Schnelle, H.~Borrmann, Y.~Grin, C.~Felser, and B.~Yan,
    Nat. Phys. {\bf 11}, 645 (2015).

	\bibitem{Wang-Zheng:2015} Z.~Wang, Y.~Zheng, Z.~Shen, Y.~Lu, H.~Fang, F.~Sheng, Y.~Zhou, X.~Yang, Y.~Li, C.~Feng, and Z.-A.~Xu,
    Phys. Rev. B {\bf 93}, 121112(R) (2016).

	\bibitem{Zhang-Xu:2015} C.-L.~Zhang, S.-Y.~Xu, I.~Belopolski, Z.~Yuan, Z.~Lin, B.~Tong, G.~Bian, N.~Alidoust, C.-C.~Lee, S.-M.~Huang, T.-R.~Chang, G.~Chang, C.-H.~Hsu, H.-T.~Jeng, M.~Neupane, D.~S.~Sanchez, H.~Zheng, J.~Wang, H.~Lin, C.~Zhang, H.-Z.~Lu, S.-Q.~Shen, T.~Neupert, M.~Z.~Hasan, and S.~Jia,
    Nat. Commun. {\bf 7}, 10735 (2016).

	\bibitem{Hasan-Huang:2017-Rev} M.~Z.~Hasan, S.-Y.~Xu, I.~Belopolski, and C.-M.~Huang,
 Ann. Rev. Cond. Mat. Phys. {\bf 8}, 289 (2017).

	\bibitem{Yan-Felser:2017-Rev} B.~Yan and C.~Felser,
 Ann. Rev. Cond. Mat. Phys. {\bf 8}, 337 (2017).

	\bibitem{Armitage-Vishwanath:2017-Rev} N.~P.~Armitage, E.~J.~Mele, and A.~Vishwanath,
 Rev. Mod. Phys. {\bf 90}, 15001 (2018).

	\bibitem{Berry:1984} M.~V.~Berry, Proc. R. Soc. London Ser. A {\bf 392}, 45 (1984).

	\bibitem{Nielsen-Ninomiya-1} H.~B.~Nielsen and M.~Ninomiya, Nucl. Phys. B {\bf 185}, 20 (1981).

	\bibitem{Nielsen-Ninomiya-2} H.~B.~Nielsen and M.~Ninomiya, Nucl. Phys. B {\bf 193}, 173 (1981).

	\bibitem{Aji} V.~Aji, Phys. Rev. B {\bf 85}, 241101 (2012).

	\bibitem{Haldane} F.~D.~M.~Haldane, arXiv:1401.0529. 

	\bibitem{Xu-Hasan:2015} S.-Y.~Xu, C.~Liu, S.~K.~Kushwaha, R.~Sankar, J.~W.~Krizan, I.~Belopolski, M.~Neupane, G.~Bian, N.~Alidoust, T.-R.~Chang, H.-T.~Jeng, C.-Y.~Huang, W.-F.~Tsai, H.~Lin, P.~P.~Shibayev, F.-C.~Chou, R.~J.~Cava, M.Z.~Hasan,
    Science {\bf 347}, 294 (2015). 

	\bibitem{Potter-Vishwanath:2014} A.~C.~Potter, I.~Kimchi, and A.~Vishwanath
    Nat. Commun. {\bf 5}, 5161 (2014). 

	\bibitem{Moll:2016} P.~J.~W.~Moll, N.~L.~Nair, T.~Helm, A.~C.~Potter, I.~Kimchi, A.~Vishwanath, and J.~G.~Analytis,
 Nature {\bf 535}, 266 (2016). 

	\bibitem{Yang-Nagaosa:2014} B.-J.~Yang and N.~Nagaosa,
    Nat. Commun. {\bf 5}, 4898 (2014).

	\bibitem{Gorbar:2014sja} E.~V.~Gorbar, V.~A.~Miransky, I.~A.~Shovkovy, and P.~O.~Sukhachov,
  Phys. Rev. B {\bf 91}, 121101 (2015).  

	\bibitem{Gorbar:2015waa} E.~V.~Gorbar, V.~A.~Miransky, I.~A.~Shovkovy, and P.~O.~Sukhachov,
  Phys. Rev. B {\bf 91}, 235138 (2015).  

	\bibitem{Yang-Furusaki:2015} B.-J.~Yang, T.~Morimoto, and A.~Furusaki,
    Phys. Rev. B {\bf 92}, 165120 (2015).

	\bibitem{Fang-Fu:2015} C.~Fang, Y.~Chen, H.-Y.~Kee, and L.~Fu,
    Phys. Rev. B {\bf 92}, 081201 (2015).

	\bibitem{Kobayashi-Sato:2015} S. Kobayashi and M. Sato,
    Phys. Rev. Lett. {\bf 115}, 187001 (2015).

	\bibitem{Burkov-Kim:2015} A. A. Burkov and Y. B. Kim,
    Phys. Rev. Lett. {\bf 117}, 136602 (2016).

	\bibitem{Rogatko:2018moa} M.~Rogatko and K.~I.~Wysokinski,
  arXiv:1804.02202.

	\bibitem{Wilczek:1984dh} F.~Wilczek and A.~Zee,
  Phys. Rev. Lett. {\bf 52}, 2111 (1984).

	\bibitem{Son:2012wh} D.~T.~Son and N.~Yamamoto,
  Phys. Rev. Lett. {\bf 109}, 181602 (2012).

	\bibitem{Stephanov:2012ki} M.~A.~Stephanov and Y.~Yin,
  Phys. Rev. Lett. {\bf 109}, 162001 (2012).  

	\bibitem{Chen:2014cla} J.~Y.~Chen, D.~T.~Son, M.~A.~Stephanov, H.~U.~Yee, and Y.~Yin,
  Phys. Rev. Lett. {\bf 113}, 182302 (2014). 

	\bibitem{Manuel:2014dza} C.~Manuel and J.~M.~Torres-Rincon,
  Phys. Rev. D {\bf 90}, 076007 (2014). 

	\bibitem{Shindou:2005vfm} R.~Shindou and K.~I.~Imura,
  Nucl. Phys. B {\bf 720}, 399 (2005).

	\bibitem{Culcer-Niu:2005} D.~Culcer, Y.~Yao, and Q.~Niu,
    Phys. Rev. B {\bf 72}, 085110 (2005).

	\bibitem{Chang:2008zza} M.~C.~Chang and Q.~Niu,
  J. Phys. Condens. Matter {\bf 20}, 193202 (2008).

	\bibitem{Xiao:2009rm} D.~Xiao, M.~C.~Chang, and Q.~Niu,
  Rev. Mod. Phys. {\bf 82}, 1959 (2010). 

	\bibitem{Gorbar:2017dtp} E.~V.~Gorbar, V.~A.~Miransky, I.~A.~Shovkovy, and P.~O.~Sukhachov,
  Phys. Rev. B {\bf 95}, 241114 (2017). 

	\bibitem{Liang-Ong:2015} T.~Liang, Q.~Gibson, M.~N.~Ali, M.~Liu, R.~J.~Cava, and N.~P.~Ong,
    Nat. Mater. {\bf 14}, 280 (2015). 

\end{thebibliography}
\end{document}